\documentclass{elsart}
\usepackage{graphicx}
\usepackage{amssymb}
\newcommand{\bfa}{{\bf a}}
\newcommand{\bfb}{{\bf b}}

\newcommand{\bfh}{{\bf h}}
\newcommand{\bfz}{{\bf z}}
\newcommand{\LL}{{\mathcal L}_0}

\newtheorem{theorem}{Theorem}[section]

\newtheorem{lemma}{Lemma}[section]
\newtheorem{corollary}{Corollary}[section]

\theoremstyle{definition} \theoremstyle{statement}

\newtheorem{statement}{Statement}[section]
\newtheorem{definition}{Definition}[section]

\newtheorem{conjecture}{Conjecture}[section]

\theoremstyle{remark}

\begin{document}
\begin{frontmatter}


\title{On classification of intrinsic localized modes for the Discrete Nonlinear
       Schr\"{o}dinger Equation}
\author[Madrid]{G. L. Alfimov \thanksref{Zelen}},
\thanks[Zelen]{On sabbatical leave from Lukin Institute
       of Physical Problems, 103460, Zelenograd, Moscow, Russia}
\ead{alfimov@nonlin.msk.ru}
\author[Lisbon1]{V. A. Brazhnyi\corauthref{cor}},
\corauth[cor]{Corresponding author} \ead{brazhnyi@cii.fc.ul.pt}
\author[Lisbon2]{V. V. Konotop}
\ead{konotop@cii.fc.ul.pt}

\address[Madrid]{Departamento de Matem\'atica Aplicada, Facultad de
  Informatica, Universidad Complutense, 28040, Madrid, Spain}
\address[Lisbon1]{Centro de F\'{\i}sica Te\'orica e
Computacional,Universidade de Lisboa, Av.Prof. Gama Pinto 2, 1649-003 Lisboa,
Portugal}
\address[Lisbon2]{Centro de F\'{\i}sica Te\'orica e
Computacional and Departamento de F\'{\i}sica, Universidade de
Lisboa, Av.Prof. Gama Pinto 2, 1649-003 Lisboa,
Portugal}
\begin{abstract}
We consider localized modes (discrete breathers) of the discrete
nonlinear Schr\"{o}\-din\-ger equation
$i\frac{d\psi_n}{dt}=\psi_{n+1}+\psi_{n-1}-2\psi_n+\sigma|\psi_n|^2\psi_n$,
$\sigma=\pm1$, $n\in \mathbb{Z}$. We study the diversity of the
steady-state solutions of the form $\psi_n(t)=e^{i\omega t}v_n$
and the intervals of the frequency, $\omega$, of their existence.
The base for the analysis is provided by the anticontinuous limit
($\omega$ negative and large enough) where all the solutions can
be coded by the sequences of three symbols ``-", ``0" and ``+".
Using dynamical systems approach we show that this coding is valid
for $\omega<\omega^*\approx -3.4533$ and the point $\omega^*$ is a
point of accumulation of saddle-node bifurcations. Also we study
other bifurcations of intrinsic localized modes which take place
for $\omega>\omega^*$ and give the complete table of them for the
solutions with codes consisting of less than four symbols.
\end{abstract}


\end{frontmatter}

\section{Introduction}\label{Int}

The discrete nonlinear Schr\"{o}dinger equation (DNLS)
\begin{equation}
i\frac{d\psi _{n}}{dt}=\Delta _{2}\psi _{n}+\sigma \mid \psi _{n}
\mid ^{2}\psi _{n},
\label{DNLS_t}
\end{equation}
where $\psi _{n}\equiv \psi _{n}(t)$, $\Delta _{2}\psi
_{n}\equiv\psi _{n+1}-2\psi _{n}+\psi _{n-1}$, and $\sigma=\pm 1$,
is one of the basic lattice models which appeared in last decades
in various contexts of physics and biology and has been
intensively studied (for review see e.g.
\cite{Surv_Tsir,Surv_Kevr,Surv_Flach,Eilbeck}). One of the recent
applications of (\ref{DNLS_t}) is related to the theory of
Bose-Einstein condensates in optical lattices, where the
mean-field Gross-Pitaevskii equation with a periodic potential is
reduced to (\ref{DNLS}) within the framework of the so-called
tight-binding approximation (see e.g. \cite{Bose} and references
therein). A particular attention has been paid to specific
solutions of (\ref{DNLS_t}) which are spatially localized and
periodic in time and called {\it intrinsic localized modes} or,
alternatively, {\it discrete breathers}. For numerous applications
of model (\ref{DNLS_t}) it is desirable to have at hands a
practical guide where possible forms of intrinsic localized modes
were described together with the information about their
stability. The aim of this paper is to make a step toward the
compilation of this guide.

The simplest class of the discrete breathers can be obtained using
the steady-state ansatz $\psi _{n}(t)=e^{i\omega t}u_{n}$. Then
the stationary amplitudes $u_{n}$, $n=0,\pm 1,\pm 2...$ satisfy
the lattice equation
\begin{equation}
\Delta _{2}u_{n}+\omega u_{n}+\sigma \mid u _{n}\mid ^{2}u _{n}=0
\label{DNLS}
\end{equation}
and the condition of spatial localization
\begin{equation}
u_{n}\rightarrow 0\quad \mbox{as}\quad n\rightarrow \pm \infty.
\label{Local}
\end{equation}

Eq.(\ref{DNLS}) is quite general and appears independently in other
applications, for example, in polaron theory as the equation for
stationary localized solutions (i.e. polarons) derived on the
basis of the semiclassical Holstein model \cite{prb58}.
Generically, $\sigma $ can be both positive and negative, $\omega
$ can be any real value and the amplitudes $u_{n}$, $n\in
\mathbb{Z}$ can be complex. However, for localized solutions
(\ref{Local}) the analysis can be simplified by taking into
account the following points:

(a) One can show \cite{Surv_Tsir} that all the solutions of Eq.(\ref{DNLS}) which satisfy the localization condition
(\ref{Local}) have the form $u_{n}=v_{n}e^{i\theta }$,  where
$n\in \mathbb{Z}$, $v_{n}$ and $\theta $ are real.

(b) The localization condition (\ref{Local}) implies that $\omega
$ does not lie in the phonon band defined by $0<\omega <4$.
Solutions for $\omega <0$ and for $\omega >4$ are connected by the
following staggering transform: if discrete function $\{u_{n}\}$,
$n\in \mathbb{Z}$ is a solution of (\ref{DNLS}) for $\sigma
=\sigma _{0}$, $\omega =\omega _{0}<0$ then $\{(-1)^{n}u_{n}\}$ is
a solution of (\ref{DNLS}) for $\sigma =-\sigma _{0}$ and $\omega
=4-\omega _{0}$.

(c) One can show \cite{Morg} that for $\omega <0$ and $\sigma =-1$
(or, correspondingly, for $\omega >4$ and $\sigma =1$) all the
solutions of (\ref {DNLS}) except the trivial zero solution are
unbounded i.e. $\mid u_{n}\mid \rightarrow \infty $ as $n$ tends
either to $+\infty $ or to $-\infty $.

Thus, for the study of {\it localized} solutions
Eq.(\ref{DNLS}) can be replaced by
\begin{equation}
\Delta _{2}u_{n}+\omega u_{n}+u _{n}^{3}=0,  \label{R_DNLS}
\end{equation}
where the discrete function $\{u_{n}\}$, $n\in \mathbb{Z}$ is real
and $\omega <0$.

Solutions of (\ref{R_DNLS}) have been widely discussed in physical
and mathematical literature.   Apart from finding important
particular examples, such as the Sievers-Takeno mode \cite{ST},
the Page mode \cite {Page}, the twisted modes
\cite{Spatschek96,Darm}, some authors have addressed the problem
of the existence and stability of more complex objects
\cite{Spatschek96}. Recently, the analysis of general structure of
the set of localized solutions of Eq.(\ref{R_DNLS}) was proposed
in \cite{Bount2,Bount3}. In particular, Ref. \cite{Bount2} is
devoted to classification of symmetric localized solutions of
(\ref{R_DNLS}) while in Ref. \cite{Bount3} a quite general
approach for the study of both symmetric and non-symmetric
solutions of Eq.(\ref{R_DNLS}) was elaborated. Both these papers
offer a coding for the solutions of Eq.(\ref{R_DNLS}).  Some
discussion of these results in comparison with our ones can be
found below in Sect.\ref{Concl}.

In the present paper we study different types of localized modes
which can coexist at a given parameter $\omega$. The general
scheme of our approach is as follows. For large values of
$|\omega|$ we describe solutions in terms of coding sequences
using {\it anticontinuous limit approach} \cite{McKay1994}. To
this end we rewrite Eq.(\ref{R_DNLS}) in the form
\begin{equation}
\alpha \Delta _{2}v_{n}-v_{n}+v_{n}^{3}=0,\quad \alpha \equiv
-1/\omega. \label{AC_L}
\end{equation}
where $u_{n}=\sqrt{-\omega }v_{n}$, $n\in \mathbb{Z}$. Next, we
calculate the boundary $\omega^*$ until which this coding works.
Finally, we study bifurcations of the solutions, describing them
in terms of their coding sequences. This step is implemented using
the representation of Eq.(\ref{R_DNLS}) as a map in
$\mathbb{R}^2$. In this part our work was influenced by the paper
\cite{Meiss1999} where a similar problem was analyzed for the
H\'enon map. The information about the bifurcations of the
simplest solutions (the solutions which have codes of length
smaller then 4) is collected in Table \ref{TabBif}.

The paper is organized as follows. In Section \ref{Ant_lim} we
describe the coding of the solutions which comes from
anticontinuous limit and give an estimation for the interval where
it is valid. In Section \ref{Dyn_sys} we study the problem from
the viewpoint of dynamical systems introducing a 2D map $T$
associated with the discrete equation (\ref{AC_L}). In these terms
the problem is reduced to the analysis of the intersection points
of stable $W_s$ and unstable $W_u$ manifolds of zero hyperbolic
fixed point of $T$. We establish the ordering of points of
intersection on these manifolds. Using this information in Section
\ref{Bif} we describe the homoclinic bifurcations. Section
\ref{Concl} contains a summary and discussion of the results.

\section{Anticontinuous limit}\label{Ant_lim}

The anticontinuous limit (ACL) corresponds to the zero value of
the coupling constant $\alpha$ in Eq.(\ref{AC_L}). When $\alpha=0$
the amplitudes of the sites $v_{n}$ are independent from each
other and can acquire one of the three values as  follows: $0$ and
$\pm 1$. Thus any solution $\{v_{n}\}$ can be coded by an infinite
sequence $\{s_{n}\}$, $n\in \mathbb{Z}$, where $s_{n}$ is one of
the symbols, $-$, $0$, and $+$, corresponding to $v_{n}=-1$,
$v_{n}=0$ and  $v_{n}=1$. Localized solutions in this case can be
identified with those of the sequences which start and end with
infinite number of consecutive zeros and contain a finite number
of nonzero elements (an example is $\{\ldots
,0,0,+,-,+,0,0,\ldots\}$). We denote the set of all these
sequences by $\LL$. It is convenient to define the code and the
length of the sequences from $\LL$.

\begin{definition}
A {\it code} of a sequence $\{\eta\}\in \LL$ is the maximal
subsequence of $\{\eta\}$ which has  nonzero first and last
elements.
\end{definition}
\begin{definition}
A {\it length} of a sequence $\{\eta\}\in \LL$ is a
number of symbols in its code.
\end{definition}
\begin{definition}
A {\it word} is a sequence of the symbols $\{-,0,+\}$.  A word
which consists of a finite number of symbols is called a finite
word.
\end{definition}

The above mentioned example
$\{\ldots ,0,0,+,-,+,0,0,\ldots\}$ has the code $\{+-+\}$ and its
length is equal to 3.

It is known \cite{McKay1994} that all solutions which exist in the
limit $\alpha =0$ can be continued by $\alpha $ up to some
positive value $\alpha ^*$. Moreover, the solutions  corresponding
to the sequences from $\LL$ remain localized under this
continuation and their decay as $n\rightarrow \pm \infty $ is
exponential.  A basis for the above statements due to
\cite{McKay1994} is given  in Appendix~\ref{App0}.

 In literature there are various bounds
for the limit value $\alpha^*$.  Application of Theorem 9 of the
paper \cite{Meiss2000} to Eq.(\ref{AC_L}) yields
$\alpha^*>1/(10+4\sqrt{2})\approx 0.0639$. The theorem from
Section 3.3 of \cite{Surv_Tsir} states that
$\alpha^*>1/6\sqrt{5}\approx 0.0745$. Below we show that
$\alpha^*>(3\sqrt{3}-1)/52\approx 0.0807$ by making use of the
following statement


\begin{lemma}
\label{Lemma_Ant_lim_1}
For each bounded solution
${\bf v}=\{v_n\}$, $n\in\mathbb{Z}$, of Eq.(\ref{AC_L}) the
following estimation is valid:
\begin{equation}
\sup_{n\in\mathbb{Z}} |v_n-v_n^3|\leq 4\alpha\sqrt{1+4\alpha}. \label{InEqLemma1}
\end{equation}
\end{lemma}
{\it Proof.}
Suppose that ${\bf v}$
is a solution of (\ref{AC_L}), (\ref{Local})  and there exists
$\rho_0>4\alpha\sqrt{1+4\alpha}$ and $n_0$ such that
$|v_{n_0}-v_{n_0}^3|\geq \rho_0$. This implies that
$|v_{n_0+1}-2v_{n_0}+v_{n_0-1}|>\rho_0/\alpha$ and, consequently,
there exists $m\in\{n_0-1,n_0,n_0+1\}$ such that
$|v_m|>\rho_0/4\alpha$ and
\begin{eqnarray*}
|v_m-v_m^3|>\left(\frac{\rho_0}{4\alpha}\right)^3-
\left(\frac{\rho_0}{4\alpha}\right)\equiv \rho_1>4\alpha.
\end{eqnarray*}
Repeating the same procedure one can construct the sequence
$\rho_0,\rho_1,\rho_2,...$, such that
\begin{eqnarray*}
\rho_{k+1}=\left(\frac{\rho_k}{4\alpha}\right)^3-
\left(\frac{\rho_k}{4\alpha}\right).
\end{eqnarray*}
By simple graphical arguments one can show that if
$\rho_0>4\alpha\sqrt{1+4\alpha}$ then $\rho_k\to\infty$
monotonically as $k\to\infty$. Since $\sup_n|v_n-v_n^3|>\rho_k$
for any $k=0,1,2...$ we arrive to the contradiction which
proves the relation (\ref{InEqLemma1}) $\blacksquare$.

Let $\alpha$ be fixed and ${\bf v}=\{v_n\}$, $n\in \mathbb{Z}$ be
a bounded solution of (\ref{AC_L}). Define the function
\begin{equation}
M(\alpha)\equiv \sup_{\bf v}\sup_{n\in\mathbb{Z}} \min
\{|v_n|,|v_n-1|,|v_n+1|\}. \label{DefM}
\end{equation}
where the first supremum is taken over all bounded solutions of
(\ref{AC_L}).  Lemma \ref{Lemma_Ant_lim_1} implies


\begin{corollary}
\label{Cor_Ant_lim_1} The function $M(\alpha)$ is correctly
defined and $M(\alpha)\to 0$ when $\alpha\to 0$.
\end{corollary}

Using Lemma \ref{Lemma_Ant_lim_1} one can prove also the following
statement:


\begin{theorem}
\label{Theor_Ant_lim_2}
For some interval
$0<\alpha<\alpha^*$, where $\alpha^*\geq
\alpha_0=(3\sqrt{3}-1)/52$ there exists one-to-one correspondence
between the set of {\it all} localized solutions of
Eq.(\ref{AC_L}) and the set of all sequences from $\LL$.
\end{theorem}

The proof can be found in Appendix \ref{AppA}. Reformulating this
statement for initial equation (\ref{R_DNLS}) results in the
following theorem

\begin{theorem}
\label{Theor_Ant_lim_3}
For some interval $-\infty<\omega<\omega^*$, where
$\omega^*\geq\omega_0=-52/(3\sqrt{3}-1)\approx -12.3923$ there
exists one-to-one correspondence between the set of all localized
solutions of Eq.(\ref{R_DNLS}) and the set of all sequences from
$\LL$. As $\omega \to -\infty$ the amplitudes of the sites denoted
by $\left\{\pm\right\}$ tend to $\pm\sqrt{-\omega }$, whereas the
amplitudes of the sites denoted by $\left\{ 0\right\} $ tend to
$0$.
\end{theorem}

\begin{figure}
\includegraphics[scale=0.5]{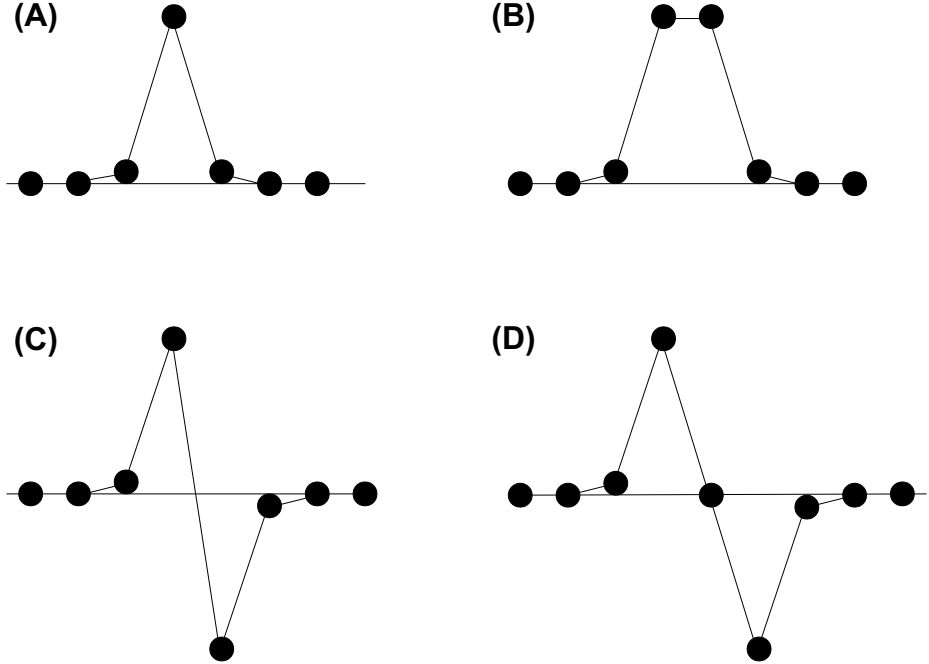}
\caption {Some simplest modes of Eq.(\ref{AC_L}): (a) the
Sievers-Takeno mode (the code $\{+\}$); (b) the Page mode (the code
$\{++\}$); (c,d) the twisted modes (the codes $\{+-\}$ and
$\{+0-\}$).} \label{Modes}
\end{figure}

Theorem \ref{Theor_Ant_lim_2} establishes a correspondence between
the localized solutions of (\ref{AC_L}) and the codes of the
sequences from $\LL$ for $\alpha$ ``close '' to anticontinuous
limit. We call this correspondence {\it ACL coding.} As  examples
of such coding we mention (see Fig.\ref{Modes}) that the
Sievers-Takeno mode has the code $\{+\}$ (or $\{-\}$), the Page
mode can be coded by $\{++\}$ (or $\{--\}$) and the twisted modes
of two various types have the codes $\{+-\}$ (or $\{-+\}$) and
$\{+0-\}$ (or $\{-0+\}$), correspondingly.

The estimation for $\alpha^*$ given above is not optimal. In the
Section \ref{Bif} we calculate $\alpha^*$ numerically. For this we
use the dynamical system approach.

\section{The dynamical system approach}\label{Dyn_sys}

\subsection{Basic notations and some auxiliary
statements}\label{BasNot}

It is convenient  to present the real-valued second-order
difference equation (\ref{AC_L}) in a form of a two-dimensional
area-preserving map $T$ \cite{Surv_Tsir,Surv_Flach,pre52,pre54}
\begin{equation}
T:\left\{
\begin{array}{l}
\tilde{x}=y;\\
\tilde{y}=(2+\frac{1}{\alpha}-\frac{1}{\alpha}y^2)y-x.
\end{array}\right.
\label{InvMap}
\end{equation}
The inverse map $T^{-1}$ is given by
\begin{equation}
T^{-1}:\left\{
\begin{array}{l}
x=(2+\frac{1}{\alpha}-\frac{1}{\alpha}\tilde{x}^2)\tilde{x}-\tilde{y};\\
y=\tilde{x}.
\end{array}\right.
\label{Map}
\end{equation}
In what follows we call a point $\bfb=T\bfa\in\mathbb{R}^2$ the
{\it $T$-image} of the point $\bfa\in\mathbb{R}^2$. The $T$-image
of a curve in $\mathbb{R}^2$ is defined as a set consisting of
$T$-images of all points of the curve.

The map $T$ is associated with Eq.(\ref{AC_L}). Specifically, if
${\bf a} =(x_0,y_0)\in \mathbb{R}^2$ is an arbitrary point of the
plane, then the bi-infinite vector formed by $x$-coordinates of
points $T^n{\bf a}$, $n\in \mathbb{Z}$ satisfies Eq.(\ref{AC_L}).
The map $T$ admits the following inversions
\begin{eqnarray}
T^{-1}\left(
\begin{array}{l} x\\ y \end{array}
\right)&=& \left(
\begin{array}{rr} 0 & 1\\1 & 0 \end{array}
\right) T \left(
\begin{array}{l} y\\ x \end{array}
\right), \label{Inv1}\\[0.2cm]
T\left(
\begin{array}{l} -x\\ -y \end{array}
\right)&=& \left(
\begin{array}{rr} -1 & 0\\0 & -1 \end{array}
\right) T \left(
\begin{array}{l} x\\ y \end{array}
\right). \label{Inv2}
\end{eqnarray}

For $\alpha>0$ and $\alpha<-1/4$ the point ${\mathcal O}=(0,0)$ is
a hyperbolic fixed  point (see e.g. \cite{Surv_Tsir}) of $T$. It
possesses invariant stable, $W_s$, and unstable, $W_u$, manifolds.
\begin{figure}
\centerline{ \scalebox{0.8}[0.8]{ \rotatebox{-90}
{\includegraphics{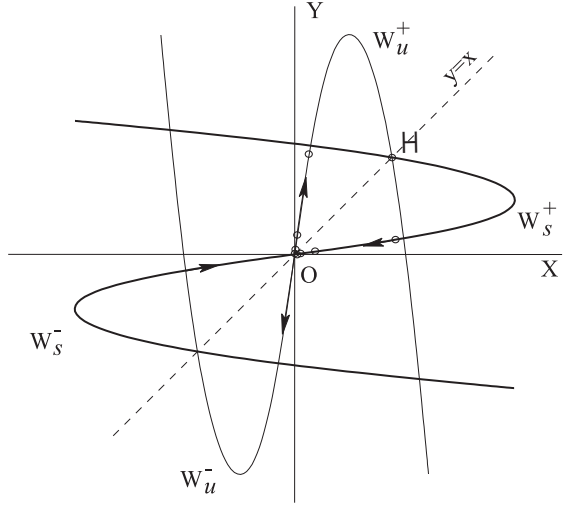}}} } \caption {An example of the
branches of stable $W_s^{\pm}$ and unstable $W_u^\pm$ manifolds of
the hyperbolic fixed point ${\mathcal O}.$ The point $H$, the
intersection of $W_s^+$ and $W_u^+$, is a homoclinic point. The
circles mark the homoclinic orbit associated with it. This orbit
corresponds to the Page mode for Eq.(\ref{AC_L}).}
\label{Manifolds}
\end{figure}
Both $W_u$ and $W_s$ consist of two branches (see
Fig.\ref{Manifolds}): $W_s=W_s^+\cup W_s^-$ and $W_u=W_u^+\cup
W_u^-$. We denote by $W_u^+$ the branch of manifold $W_u$ which
goes out from the hyperbolic fixed point ${\mathcal O}$ keeping
$y$ positive. Correspondingly, $W_s^+$ is the branch of manifold
$W_s$ which enters the hyperbolic fixed point ${\mathcal O}$ with
positive $x$. The branches are connected by the symmetries:
$W_{u,s}^+$ and $W_{u,s}^-$ are symmetric with respect to
${\mathcal O}$ whereas $W^{\pm}_{s,u}$ and $W^{\pm}_{u,s}$ are
symmetric with respect to the axis $y=x$. So, any of the four
branches allows one to reconstruct three others. Some analytical
approximations for $W^{\pm}_{u,s}$ can be constructed using normal
form analysis, see \cite{pre54}.

Points of intersections of $W_s$ and $W_u$ ({\it homoclinic
points}) correspond to localized solutions of Eq.(\ref{AC_L}). As
it was shown above, when $0<\alpha<\alpha^*$ all localized
solutions can be coded by the sequences from $\LL$. The homoclinic
points also can be associated with these sequences. Each
homoclinic point has the only matching sequence from $\LL$, but
each coding sequence from $\LL$ corresponds to an infinite number
of homoclinic points connected with each other by iterations of
$T$ ({\it homoclinic orbit}). This is a natural consequence of the
translational invariance of the lattice. Due to the involutions
(\ref{Inv1}), (\ref{Inv2}) the points of intersections of
$W_s^\pm$  (and $W_u^\pm$) with the lines $y=x$ and $y=-x$ are
homoclinic points. Their codes are, correspondingly, symmetric or
antisymmetric.

Let us introduce the following definitions.

\begin{definition}
We call basic points the nine points $(0,0)$, $(-1,0)$, $(1,0)$,
$(0,1)$, $(-1,1)$, $(1,1)$,$(0,-1)$, $(-1,-1)$, $(1,-1)$ on the
plane $(x,y)$.
\end{definition}

\begin{definition}
The spot $S_{sk}$, $s,k\in \{-,0,+\}$ is a closed square on the
plane $(x,y)$ with the side $2M(\alpha)$, where $M(\alpha)$ is
given by Eq.(\ref{DefM}), centered at the basic point coded with
the symbols $s$ and $k$. The symbol $s$ is the sign of abscissa of
the basic point (or zero), the second one is the sign of its
ordinate (or zero).
\end{definition}

\begin{figure}
\centerline{\includegraphics[scale=0.8]{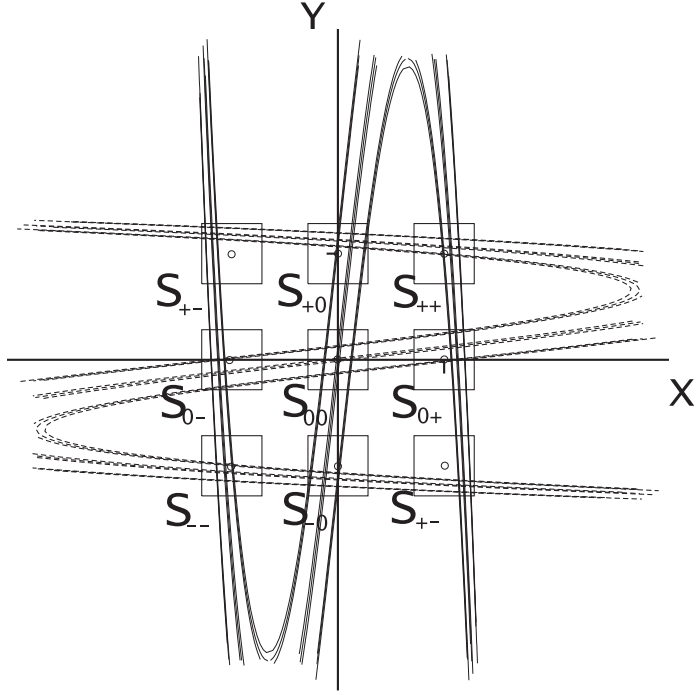}} \caption {The
spots $S_{sk}$ and stable (dashed) and unstable (solid) manifolds
of the fixed point $\mathcal O$. Small circles mark the basic
points.} \label{Spots}
\end{figure}

For example, $S_{+-}$ is the spot which is centered in the point
$(1,-1)$ (see Fig.~\ref{Spots}). It follows from
Corollary~\ref{Cor_Ant_lim_1} that
\begin{itemize}
\item[(i)]
 for $\alpha$ small enough each homoclinic point on the plane
$(x,y)$ is situated in the vicinity of one of the basic points.
When $\alpha$ tends to zero the distance between the homoclinic
point and the corresponding basic point also tends to zero;
\item[(ii)]
 all homoclinic
points which tend to a basic point as $\alpha$ goes to zero can be
covered by the spot centered at this basic point.
\item[(iii)]
if $\alpha$ is small enough the spots corresponding to different
basic points do not intersect each other.
\end{itemize}

An important fact that follows from (\ref{InvMap}) is that $T$
maps the homoclinic points from a spot $S_{sk}$, $s,k\in
\{-,0,+\}$ to points belonging to the three spots $S_{k-}$,
$S_{k0}$, $S_{k+}$ but not to any other spots.

For what follows we need the definitions of {\it segment}, {\it
initial segment} and {\it fundamental segment} of $W_{u,s}^\pm$
(see \cite{Meiss1999,GeomMeth}). They can be found in Appendix
~\ref{AppB}. Also it is useful to introduce {\it a natural
parameterization} on the branches $W_s^\pm$ and $W_u^\pm$. Let the
parameter $\tau$ be the length of the curve $W_s^+$ measured from
the fixed point $\mathcal O$. The point $\mathcal O$ corresponds
to $\tau=0$. The coordinates of the points on the branch $W_s^+$
will be denoted by $x_s^+(\tau)$ and $y_s^+(\tau)$. In the same
manner we introduce the natural parameterization on the other
branches $W_s^-$,$W_u^+$ and $W_u^-$.

Now we prove the following auxiliary statement:

\begin{lemma}
\label{Lemma_BasNot_1}
Let
$\tilde{\gamma}=(\tilde{x}(\tau),\tilde{y}(\tau))$, $\tau\in
[\tau_1,\tau_2]$ be a curve on the plane $(x,y)$ such that
\begin{itemize}
\item[(i)]
$\tilde{x}(\tau),\tilde{y}(\tau)\in C^1[\tau_1,\tau_2]$;
\item[(ii)]
for $\tau\in [\tau_1,\tau_2]$ the curve
$\tilde{\gamma}=(\tilde{x}(\tau),\tilde{y}(\tau))$ lies entirely
within one spot $S_{s_1 s_2}$, $s_{1,2}\in \{-,0,+\}$;
\item[(iii)]
for $\tau\in [\tau_1,\tau_2]$ the functions $\tilde{x}(\tau)$ and
$\tilde{y}(\tau)$ are monotonic with respect to $\tau$;
\item[(iv)]
$|\dot{\tilde{y}}(\tau)/\dot{\tilde{x}}(\tau)|<1/2$ for all
$\tau\in [\tau_1,\tau_2]$.
\end{itemize}
Let $\gamma=(x(\tau),y(\tau))$ be $T^{-1}$-image of
$\tilde{\gamma}$. Then for $\alpha$ small enough the following
statements are valid:
\begin{itemize}
\item[(a)]
for $\tau\in [\tau_1,\tau_2]$ the functions $x(\tau)$ and
$y(\tau)$ are monotonic. Moreover, if $s_2=\{0\}$ then ${\rm
sign}~ \dot{x}(\tau)={\rm sign}~ \dot{\tilde{x}}(\tau)$ and if
$s_2\ne\{0\}$ then ${\rm sign}~ \dot{x}(\tau)=-{\rm sign}~
\dot{\tilde{x}}(\tau)$;
\item[(b)]
$|\dot{y}(\tau)/\dot{x}(\tau)|<1/2$ for all $\tau\in
[\tau_1,\tau_2]$.
\end{itemize}
\end{lemma}
{\it Proof}. Consider the case when the curve
$\tilde{\gamma}$ lies entirely in one of the spots $S_{s_1 s_2}$
with $s_2=\{0\}$; the cases of $s_2=\{+\}$ and $s_2=\{-\}$ can be
analyzed in the same manner. Differentiating (\ref{Map}) one
obtains
\begin{eqnarray}
\dot{x}&=&(2+\frac{1}{\alpha}-\frac{3}{\alpha}\tilde{x}^2)\dot{\tilde{x}}
-\dot{\tilde{y}}; \label{Diff1}\\
\dot{y}&=&\dot{\tilde{x}}. \label{Diff2}
\end{eqnarray}
The side of the spot $S_{s_1 0}$ is equal to $2M(\alpha)$, so
${\tilde x}^2<4M^2(\alpha)$. This implies that for $\alpha$ small
enough the derivatives $\dot{x}(\tau)$ and $\dot{y}(\tau)$ do not
change sign and ${\rm sign}~ \dot{x}(\tau)={\rm
sign}~\dot{\tilde{x}}(\tau)$. Dividing Eq.(\ref{Diff2}) by
Eq.(\ref{Diff1}) and taking into account (iv) one obtains (b)
$\blacksquare$.

Consider now the branch $W_s^+$. Let
$\bfz=(x_s^+(\tau_0),y_s^+(\tau_0))$ be the point of the first
intersection of $W_s^+$ with the boundary of the spot $S_{00}$.
Then the initial segment $W_s^+(\mathcal{O},\bfz]$ remains
entirely within the spot $S_{00}$ and on it the following Lemma is
valid:
\begin{lemma}
\label{Lemma_BasNot_2}
For $0<\tau<\tau_0$ and $\alpha$
small enough the functions $x_s^+(\tau)$ and $y_s^+(\tau)$ are
monotone increasing and
$|\dot{y_s^+}(\tau)/\dot{x_s^+}(\tau)|<1/2$.
\end{lemma}
{\it Proof}. First, we
note that in some small neighborhood of the point $\mathcal O$ the
map $T$ can be well approximated by its linearization. This means
that there exist $\tau^*<\tau_0$ and the constants $C_{1,2}>0$
such that for $0<\tau\leq\tau^*$
\begin{eqnarray*}
|x_s^+(\tau)-\tau\cos\varphi|&<&C_1\tau^2\leq C_1{\tau^*}^2;\\
|y_s^+(\tau)-\tau\sin\varphi|&<&C_2\tau^2\leq C_2{\tau^*}^2,
\end{eqnarray*}
where $\varphi$ is the angle between a stable eigenvector of the
linearized map and positive direction of the $x$-axis and
\begin{eqnarray*}
\tan\varphi&=&\left(1+\frac1{2\alpha}\right)-
\sqrt{\left(1+\frac1{2\alpha}\right)^2-1}=O(\alpha),\quad
(\alpha\to 0).
\end{eqnarray*}
Let  $\bfz^*=(x_s^+(\tau^*),y_s^+(\tau^*))$. Then for $\alpha$ and
$\tau^*$ small enough the initial segment $W_s^+(\mathcal
O,\bfz^*]$ satisfies the conditions of Lemma \ref{Lemma_BasNot_1}.
The transformation $T^{-1}$ maps $W_s^+(\mathcal O,\bfz^*]$ to
another initial segment of $W_s^+$ for which according to the
Lemma \ref{Lemma_BasNot_1} the conditions of the same Lemma
\ref{Lemma_BasNot_1} hold also. Repeating the procedure
iteratively one concludes that for the whole initial segment
$W_s^+(\mathcal{O},\bfz]$ the statements of Lemma
\ref{Lemma_BasNot_2} are valid $\blacksquare$.

Using Lemma \ref{Lemma_BasNot_1} one can prove the following
statements.
\begin{lemma}
\label{Lemma_BasNot_3}
For $\alpha$ small enough there
exists a fundamental segment $B_s^+\in W_s^+$ such that
\begin{itemize}
\item[(i)]
$B_s^+$ intersects the spot $S_{+0}$ and does not intersect the
other spots;
\item[(ii)]
the intersection $B_s^+\cap S_{+0}$ consists of one connected
component;
\item[(iii)]
all homoclinic points which $B_s^+$ contains belong to $S_{+0}$;
\item[(iv)] if $\bfh \in B_s^+$ is a homoclinic point then for any $n$
$T^n \bfh \in S_{00}$.
\end{itemize}
\end{lemma}

\begin{figure}
\centerline{\includegraphics[scale=0.8]{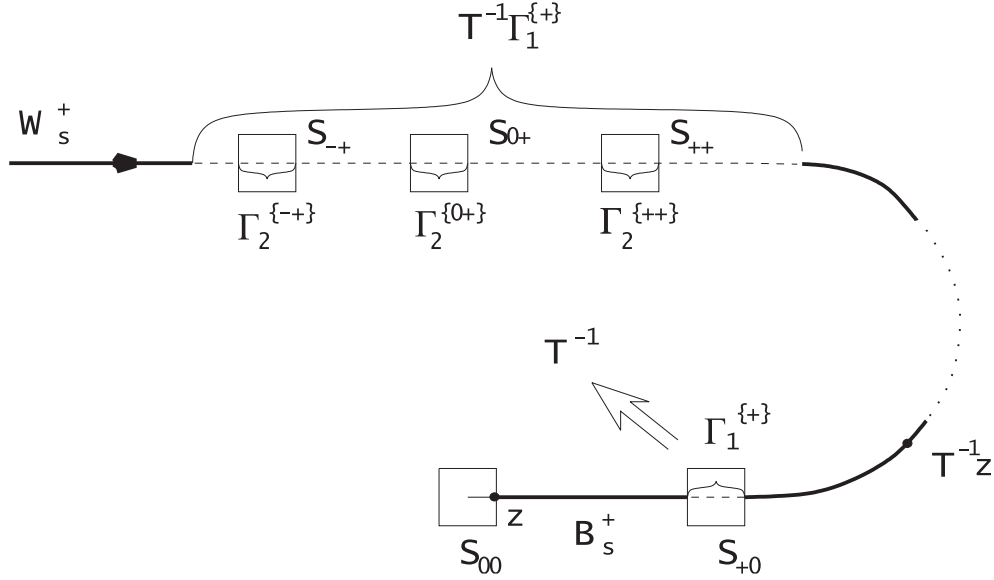}} \caption {A
basic fundamental segment $B_s^+=W_s^+({\bf z},T^{-1}{\bf z}]$.
All homoclinic points belonging to $B_s^+$ also belong to
$\Gamma^{\{+\}}_1=S_{+0}\cap B_s^+$. $T^{-1}$-images of these
points lie in $\Gamma^{\{-+\}}_2$, $\Gamma^{\{0+\}}_2$ and
$\Gamma^{\{++\}}_2$. The notations are defined in the text.} \label{Basic}
\end{figure}

{\it Proof.} Consider the fundamental segment $W_s^+(T\bfz,\bfz]$
where $\bfz=(x(\tau_0),y(\tau_0))$ is defined above (see also
Fig.~\ref{Basic}).  Evidently $W_s^+(T\bfz,\bfz]\subset S_{00}$
and for it the conditions of Lemma \ref{Lemma_BasNot_1} holds.
Define $B_s^+\equiv
T^{-1}W_s^+(T\bfz,\bfz]=W_s^+(\bfz,T^{-1}\bfz]$ (see
Fig.~\ref{Basic}). This fundamental segment can be given in a
parametric form, specifically, $B_s^+=(x_s^+(\tau),y_s^+(\tau))$,
$\tau\in(\tau_0,\tilde{\tau}]$ for some $\tilde{\tau}$ and
according to Lemma~\ref{Lemma_BasNot_1} for
$\tau_0<\tau\leq\tilde{\tau}$ the coordinate $x_s^+(\tau)$ grows
monotonically and $|\dot{y}_s^+(t)/\dot{x}_s^+(\tau)|<1/2$. Then
using simple geometrical arguments one arrives to (i) and (ii).
The point (iii) is a consequence of the fact that all homoclinic
points belong to one of the spots. The point (iii) implies also
(iv) $\blacksquare$.

By symmetry one can conclude that for $\alpha$ small enough there
exists a fundamental segment $B_s^-\subset W_s^-$ such that all
homoclinic points which belong to $B_s^-$ lie in $S_{-0}$.
Analogously, there exist the fundamental segments $B_u^+\subset
W_u^+$ and $B_u^-\subset W_u^-$, such that all homoclinic points
which belong to $B_u^\pm$ lie in $S_{0\pm}$. We call $B_s^\pm$ and
$B_u^\pm$ {\it the basic fundamental segments}.

The following statements are valid

\begin{lemma}
\label{Lemma_BasNot_4}
For $\alpha$ small enough and for any $n\in \mathbb{Z}$ there
exist the following one-to-one correspondences:
\begin{itemize}
\item[(i)]
 between the homoclinic points which
belong to the fundamental segment $T^n B_s^+$ and the set of the
sequences from ${\LL}$ which have the codes with last symbol
``$+$";
\item[(ii)]
 between the homoclinic points which
belong to the fundamental segment $T^n B_s^-$ and the set of the
sequences from ${\LL}$ which have the codes with last symbol
``$-$"
\item[(iii)]
 between the homoclinic points which
belong to the fundamental segment $T^n B_u^+$ and the set of the
sequences from ${\LL}$ which have the codes with first symbol
``$+$";
\item[(iv)]
 between the homoclinic points which
belong to the fundamental segment $T^n B_u^-$ and the set of the
sequences from ${\LL}$ which have the codes with first symbol
``$-$".
\end{itemize}
\end{lemma}

{\it Proof.} As an illustration let us consider the point (i) and
$n=0$ (the basic fundamental segment $B_s^+$ itself); all other
cases can be treated similarly. Let $h$ be a homoclinic point from
$B_s^+$. Consider the corresponding solution of Eq.(\ref{AC_L}).
Its coding sequence from $\LL$ contains the infinite block
$(+,0,0,0...)$, so the code end up with the symbol ``$+$". From
the other side, consider a sequence $L\in\LL$ which end up with
the block $(+,0,0,0...)$. If $\alpha$ is small enough, there
exists a solution of Eq.(\ref{AC_L}) which corresponds to $L$. On
the plane $(x,y)$ its image is a sequence of homoclinic points
$h_k$, $k\in\mathbb{Z}$, $T h_k=h_{k+1}$. According to the coding
sequence, there exists a number $k_0$, such that for $k>k_0$
$h_k\in S_{00}$ and $h_{k_0}\in S_{0+}$. Then $h_{k_0}\in B_s^+$
by construction of $B_s^+$ in Lemma~\ref{Lemma_BasNot_3}
$\blacksquare$.

In what follows we will label homoclinic points by their codes
saying, for example, ``homoclinic point $\{-+-\}$". This labeling
is well-defined within one fundamental segment: two points which
belong to the same fundamental segment definitely have different
codes. In what follows if the fundamental segment is not
specified, this means that the homoclinic point under
consideration belongs to $B_s^\pm$.

\subsection{Ordering on $B_s^\pm$} \label{Ord}

We say that $\bfa>_s \bfb$ ($\bfa<_s \bfb$) on the branch $W_s^\pm$ if $\bfb$ lies
closer (farther) on $W_u^\pm$ to the hyperbolic fixed point ${\mathcal O}$
than $\bfa$. The ordering on $W_u^\pm$ is defined in the same
manner.

We denote by $\{\{\xi_1\}\{\xi_2\}\}$ the word composed by two
words $\{\xi_1\}$ and $\{\xi_2\}$. For example, if
$\{\xi_1\}=\{++\}$ and $\{\xi_2\}=\{--\}$ then
$\{\{\xi_1\}\{\xi_2\}\}=\{++--\}$.

Then the following Theorem establishes the ordering on $B_s^+$:

\begin{theorem}
\label{Theor_Ord_1}
Let $\alpha$ be small enough
and $\{\xi\}$  be a word of arbitrary but finite length. Then if
the total number of symbols $\{+\}$ and $\{-\}$ in $\{\xi\}$ is
{\it odd} then for any three words
$\{\eta_1\}$,$\{\eta_2\}$,$\{\eta_3\}$ the homoclinic points on
$B_s^+$ are ordered as follows
\begin{equation} \{\{\eta_1\}-\{\xi\}+\} <_s
\{\{\eta_2\}0\{\xi\}+\}<_s \{\{\eta_3\}+\{\xi\}+\} \label{Order1}
\end{equation}
and if $\{\xi\}$ contains {\it even} number of the symbols $\{+\}$
and $\{-\}$ then
\begin{equation}
\{\{\eta_1\}+\{\xi\}+\} <_s \{\{\eta_2\}0\{\xi\}+\}<_s
\{\{\eta_3\}-\{\xi\}+\}. \label{Order2}
\end{equation}
\end{theorem}
{\it Proof.} First, let us verify this statement for the words
$\{\xi\}$ of length 1. Define $\Gamma_0^{\{+\}}\equiv
W_s^+(\mathcal O,\bfz]$ (where $\bfz$ is defined in subsection
\ref{BasNot}) and $\Gamma_1^{\{+\}}\equiv B_s^+\cap S_{+0}$ (see
Fig.~\ref{Basic}). Evidently, $T\Gamma_1^{\{+\}}\subset
\Gamma_0^{\{+\}}$. According to Lemma~\ref{Lemma_BasNot_3}
$\Gamma_1^{\{+\}}$ consists of a single connected component and
there exist $\tau_1^+$ and $\tau_2^+$ such that
$\Gamma_1^{\{+\}}=(x_s^+(\tau),y_s^+(\tau))$ for
$\tau_1^+<\tau<\tau_2^+$. Consider now $T^{-1}$-image of
$\Gamma_1^{\{+\}}$. According to Lemma~\ref{Lemma_BasNot_1}
$T^{-1}\Gamma_1^{\{+\}}=(x_s^+(\tau),y_s^+(\tau))$ where $\tau$
varies within some interval $\tilde{\tau}_1<\tau<\tilde{\tau}_2$,
the function $x_s^+(\tau)$ monotonically decreases within this
interval and
\begin{equation}
|\dot{y}_s^+(\tau)/\dot{x}_s^+(\tau)|<1/2, \quad
\tau_1^+<\tau<\tau_2^+ .\label{Angle}
\end{equation}
The homoclinic points which lie in $\Gamma_1^{\{+\}}$ are
$T$-images of homoclinic points which belong to fundamental
segment $T^{-1}B_s^+$. They are situated in the spots $S_{++}$,
$S_{0+}$ and $S_{-+}$. So, $T^{-1}\Gamma_1^{\{+\}}$ intersects
these three spots. Since $x_s^+$ decreases monotonically on
$T^{-1}\Gamma_1^{\{+\}}$ the intersections are in the order given
above (see Fig.~\ref{Basic}). It follows from (\ref{Angle}) that
$T^{-1}\Gamma_1^{\{+\}}$ does not intersect any other spot. Denote
\begin{eqnarray*}
\Gamma_2^{\{++\}}&=&T^{-1}\Gamma_1^{\{+\}}\cap S_{++},\\
\Gamma_2^{\{0+\}}&=&T^{-1}\Gamma_1^{\{+\}}\cap S_{0+},\\
\Gamma_2^{\{-+\}}&=&T^{-1}\Gamma_1^{\{+\}}\cap S_{-+}.
\end{eqnarray*}
Each of $\Gamma_2^{\{++\}}$, $\Gamma_2^{\{0+\}}$ and
$\Gamma_2^{\{-+\}}$ consists of one connected component. On the
fundamental segment $T^{-1}B_s^+$ the arc $\Gamma_2^{\{++\}}$
includes all homoclinic points corresponding to the codes
$\{\{\eta_1\}++\}$, the arc $\Gamma_2^{\{0+\}}$ includes all
homoclinic points corresponding to the codes $\{\{\eta_2\}0+\}$
and the arc $\Gamma_2^{\{-+\}}$ - all homoclinic points
corresponding to the codes $\{\{\eta_3\}-+\}$ (here $\eta_1$,
$\eta_2$ and $\eta_3$ are arbitrary words). This implies that on
$B_s^+$
\begin{eqnarray*}
\{\{\eta_1\}++\}<_s\{\{\eta_2\}0+\}<_s\{\{\eta_3\}-+\}.
\end{eqnarray*}

In general case we denote the corresponding segments by recurrence
relations
\begin{eqnarray*}
\Gamma_n^{\{+k\{\zeta\}+\}}&=&T^{-1}\Gamma_{n-1}^{\{k\{\zeta\}+\}}\cap
S_{+ k},\\
\Gamma_n^{\{0k\{\zeta\}+\}}&=&T^{-1}\Gamma_{n-1}^{\{k\{\zeta\}+\}}\cap
S_{0 k},\\
\Gamma_n^{\{-k\{\zeta\}+\}}&=&T^{-1}\Gamma_{n-1}^{\{k\{\zeta\}+\}}\cap
S_{- k},
\end{eqnarray*}
where $\{\zeta\}$ runs all the words of length $(n-2)$ and $k\in
\{-,0,+\}$. Arcs $\Gamma_n^{\{+k\{\zeta\}+\}}$, $
\Gamma_n^{\{0k\{\zeta\}+\}}$ and $\Gamma_n^{\{-k\{\zeta\}+\}}$
belong to fundamental segment $T^{-(n-1)} B^+_s$. Similarly to the
considered case using Lemma~\ref{Lemma_BasNot_1} one can conclude that

(a) $T^{-1}\Gamma_{n-1}^{\{k\{\zeta\}+\}}$ intersects the three
spots $S_{+ k}$, $S_{0 k}$ and $S_{- k}$ and does not intersect
the other spots;

(b) Each of $\Gamma_n^{\{+k\{\zeta\}+\}}$,
$\Gamma_n^{\{0k\{\zeta\}+\}}$ and $\Gamma_n^{\{-k\{\zeta\}+\}}$
consists of only one connected component, so each of these arcs
can be parameterized in the form $(x_s^+(\tau),y_s^+(\tau))$ at
appropriate interval of $\tau$.

(c) The order in which $\Gamma_n^{\{+k\{\zeta\}+\}}$,
$\Gamma_n^{\{0k\{\zeta\}+\}}$ and $\Gamma_n^{\{-k\{\zeta\}+\}}$
appear on $T^{-(n-1)} B^+_s$ coincides with the order in which
$\Gamma_{n-1}^{\{+\{\zeta\}+\}}$, $\Gamma_{n-1}^{\{0\{\zeta\}+\}}$
and $\Gamma_{n-1}^{\{-\{\zeta\}+\}}$ appear on $T^{-(n-2)} B^+_s$
if $k=\{0\}$ and is opposite to it if $k=\{+\}$ or $k=\{-\}$.

\begin{figure}
\centerline{\includegraphics[scale=0.5]{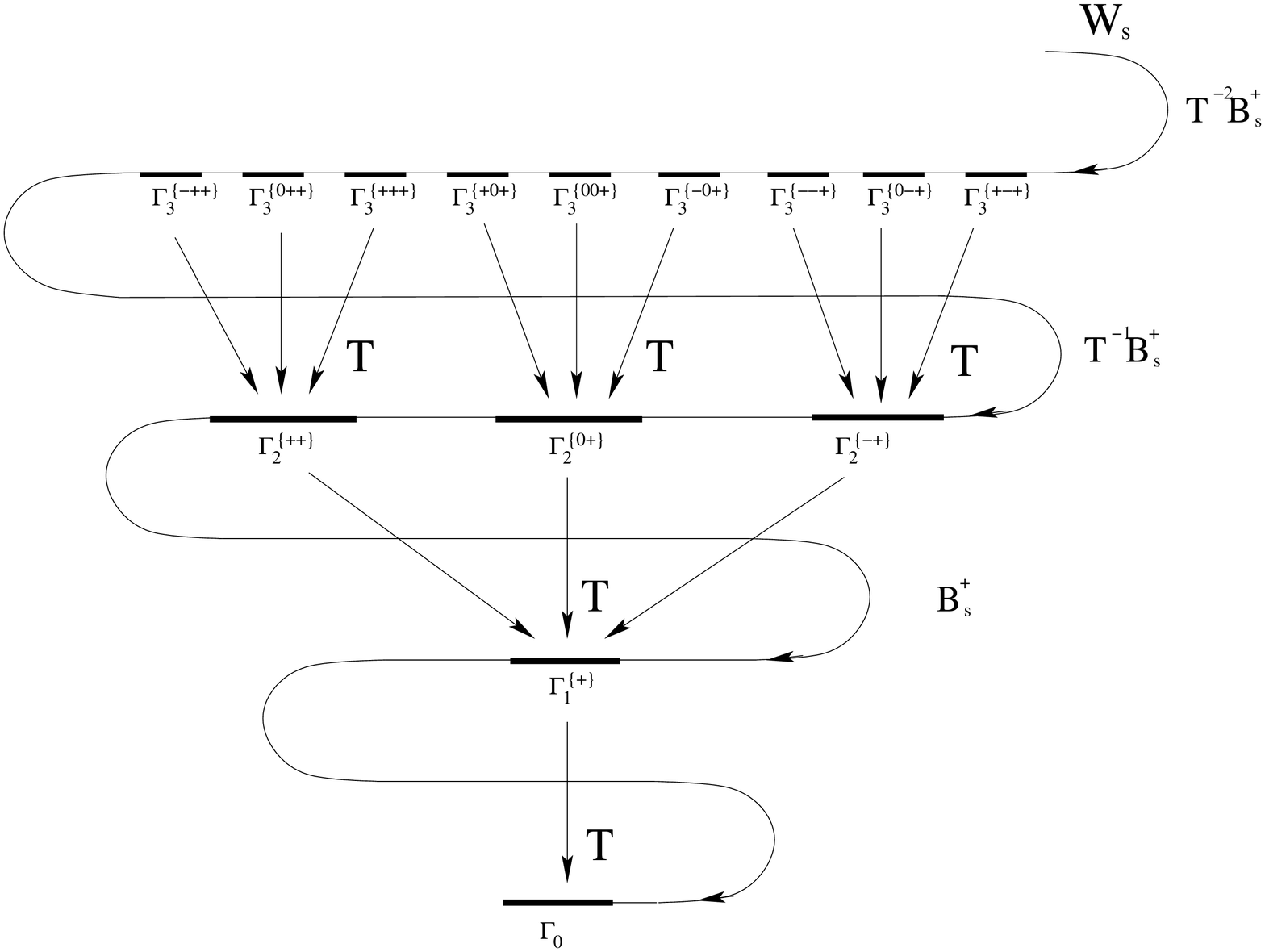}} \caption {A
schematic picture of relations between the arcs
$\Gamma_n^{\{\{\xi\}+\}}$} \label{Gammas}
\end{figure}

A schematic picture of relations between the arcs
$\Gamma_n^{\{\{\xi\}+\}}$ is shown in the Fig.~\ref{Gammas}. Taking
into account that the homoclinic points which belong to
$\Gamma_n^{\{\{\xi\}+\}}$ correspond to the codes
$\{\{\eta\}\{\xi\}+\}$ where $\eta$ is an arbitrary word we obtain
the statement of the Theorem $\blacksquare$.

{\it Remark.} Theorem~\ref{Theor_Ord_1} gives a {\it simple
recurrence algorithm} allowing one to establish the proper
ordering of homoclinic points with codes of lengths smaller than
some prescribed length on $B_s^+$. It consists in the following
steps:

\underline{Step 1}. To write the starting codes
$\{++\}$,$\{0+\}$,$\{-+\}$ (exactly in this order) and the arrow
pointed from $\{-+\}$ to $\{++\}$.

\underline{Step $n$}. Let the codes for $(n-1)$-st step be just
written.

Then one has

1. To draw arrows corresponding to each of the codes such that:

(a) the extreme left arrow is pointed in opposite direction to the
extreme left arrow on the previous step;

(b) the neighboring arrows have opposite directions.

\begin{figure}
\centerline{\includegraphics[scale=0.8]{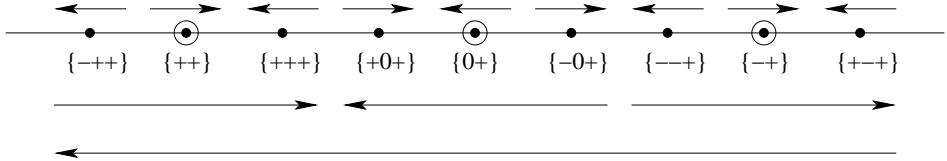}} \caption
{How to place the homoclinic points on $B_s^+$ in the proper order
(illustration to the algorithm).} \label{Algorithm}
\end{figure}

2. To generate new string of codes replacing $\{\xi\}$ by the
three codes $\{-\{\xi\}\}$, $\{0\{\xi\}\}$, $\{+\{\xi\}\}$ if the
arrow corresponding to $\{\xi\}$ is pointed to the right and
$\{+\{\xi\}\}$,$\{0\{\xi\}\}$,$\{-\{\xi\}\}$ if the arrow is
pointed to the left (see Fig.\ref{Algorithm}).


In Fig.~\ref{Algorithm} the largest arrow corresponds to the first
step of the algorithm, three smaller ones correspond to the Step 2, and nine
short arrows correspond to the third step. As a result, the picture shows
the mutual position of the homoclinic points with the
code length smaller than 3 on $B_s^+$.

\begin{figure}[tbp]
\rotatebox{90}{\scalebox{.6}{\includegraphics{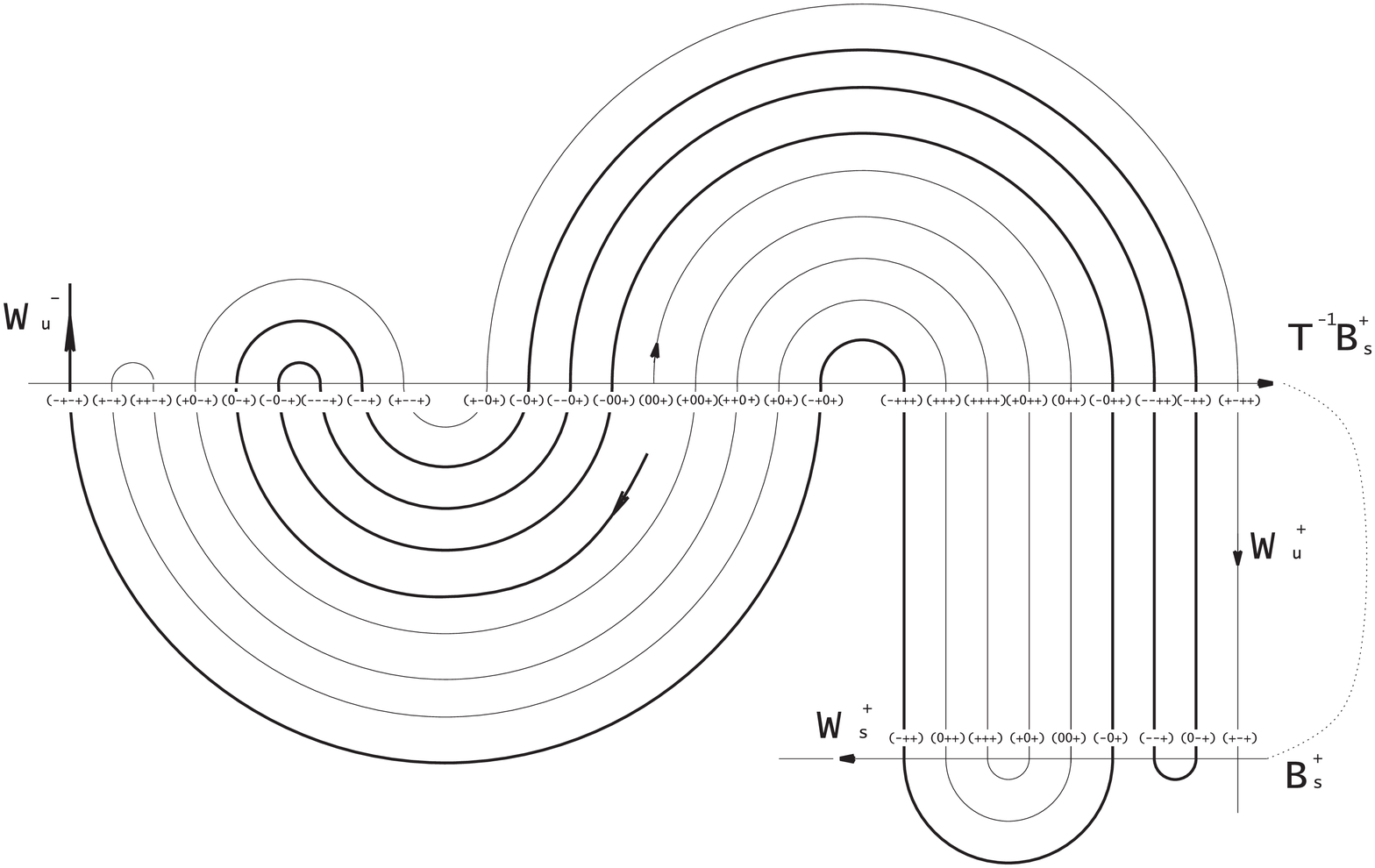}}}
\caption {Schematic representation of the intersections of $W_u^+$
(thin line) and $W_u^-$ (bold line) with the branch $W_s^+$.}
\label{Spiral}
\end{figure}
\begin{figure}
\centerline{\includegraphics[scale=0.5]{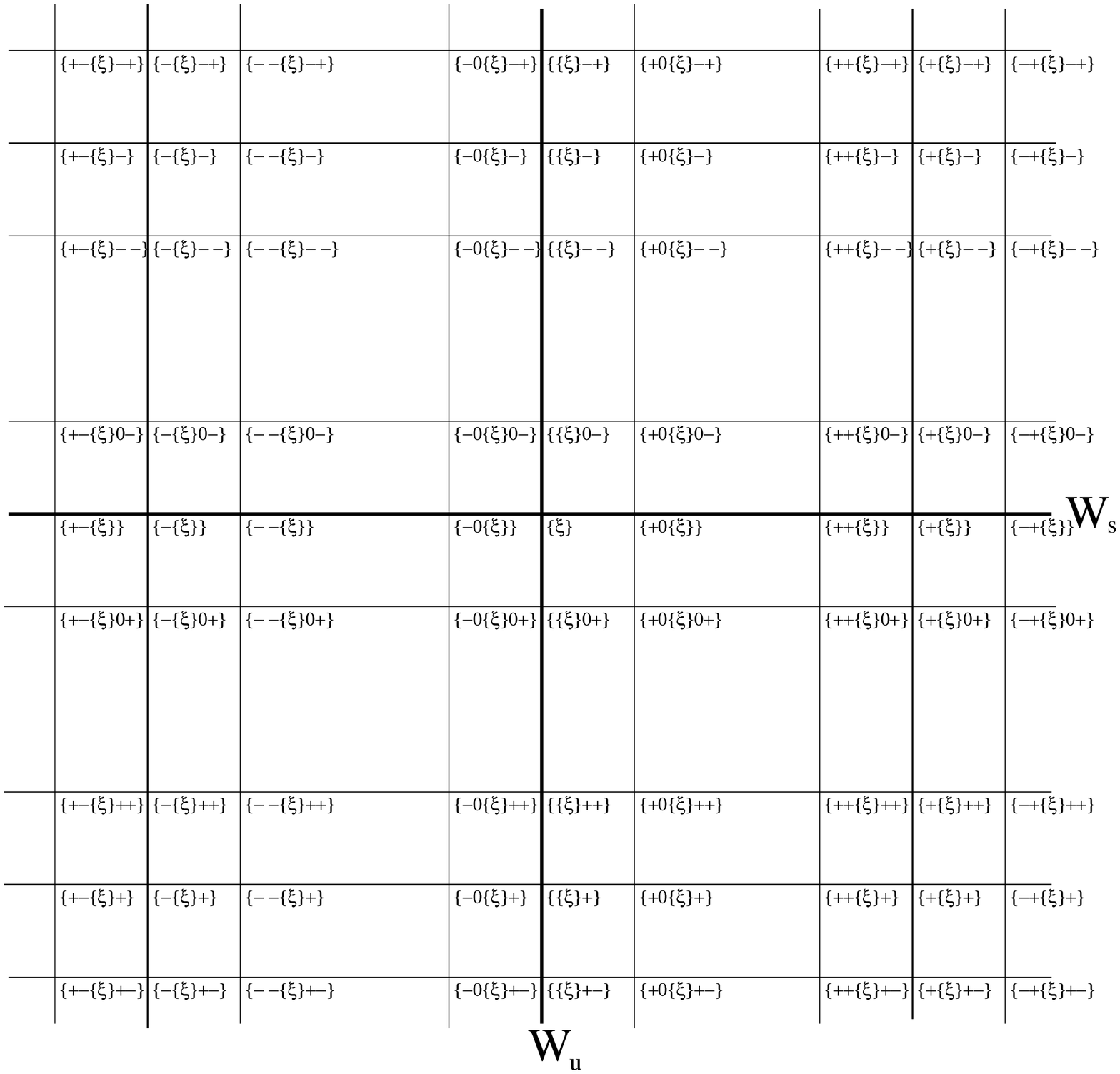}} \caption {The
homoclinic points in a neighborhood of the homoclinic point with
the code $\{\xi\}$.} \label{Grid}
\end{figure}

The same algorithm allows one to order homoclinic points on other
basic fundamental segments $B_s^-$, $B_u^+$ and $B_u^-$. By comparing
orders of homoclinic points on both stable and unstable
manifolds, it is possible to represent schematically how $W_u^+$
and $W_u^-$ intersect $W_s^+$ (see Fig.\ref{Spiral}). Also it is
possible to describe a structure of homoclinic points which are
situated in a neighborhood of a homoclinic point $\{\xi\}$ (see
Fig.\ref{Grid}).

\section{Homoclinic bifurcations} \label{Bif}

Theorem~\ref{Theor_Ant_lim_2} states that for the interval
$0<\alpha<\alpha^*$ all localized solutions of Eq.(\ref{AC_L})
are in one-to-one correspondence with the set of all sequences
from $\LL$. If $\alpha$ exceeds the critical value $\alpha^*$ this
does not take place any more and solutions can appear or disappear
as a result of some bifurcations. Since the localized solutions
correspond to the points of intersections of the stable $W_s$ and
the unstable $W_u$ manifolds of the fixed point $\mathcal O$,
these bifurcations can be found numerically by observing how the
forms of $W_s$ and $W_u$ change as $\alpha$ grows. In such a way,
each of the solutions of Eq.(\ref{AC_L}) which exists for
$0<\alpha<\alpha^*$ can be continued until some value of
$\alpha>\alpha^*$ where it dies in some bifurcation. It is
convenient to label this solution with the same code as it has in
the anticontinuous limit. In this sense, the coding of solutions
used for $0<\alpha<\alpha^*$ can be used for $\alpha>\alpha^*$
also. A problem can occur, however, if some new solutions appear
since they have no counterparts in the set of the coding
sequences.

In order to study possible bifurcations we used a numerical
procedure which allows one to construct initial segments of the
manifolds $W_u$ and $W_s$, to find their points of intersection,
and to identify them with possible codes from $\LL$. Using this
program we observed that the deformation of the manifolds $W_u$
and $W_s$ as $\alpha$ grows is quite monotonic. This results in
numerous bifurcations of death of homoclinic points. However, we
did not observe any bifurcation of birth of new homoclinic point.
This allows us to formulate the following conjecture.
\begin{figure}[tbp]
{\scalebox{.6}{\includegraphics{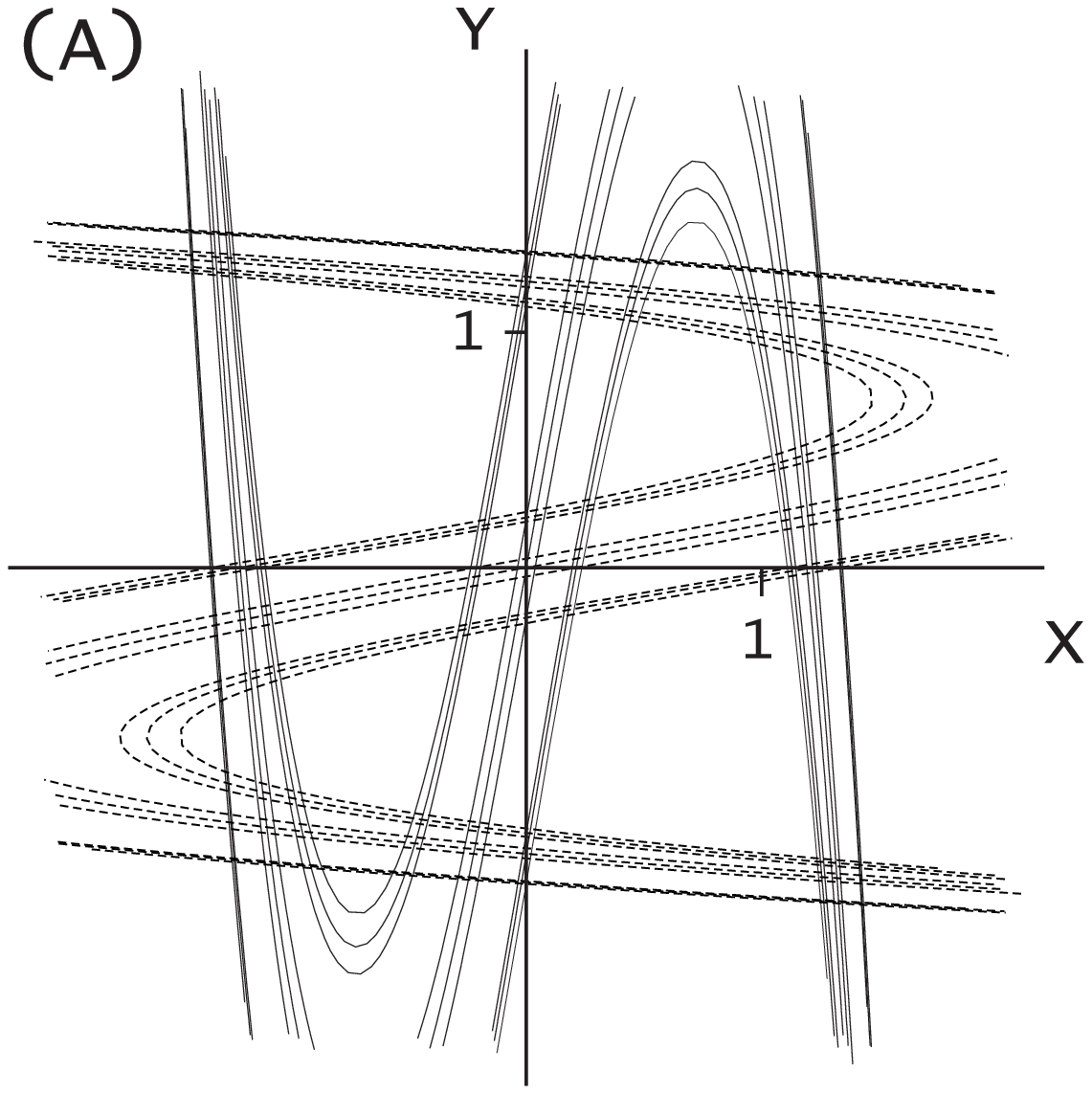}}}
{\scalebox{.6}{\includegraphics{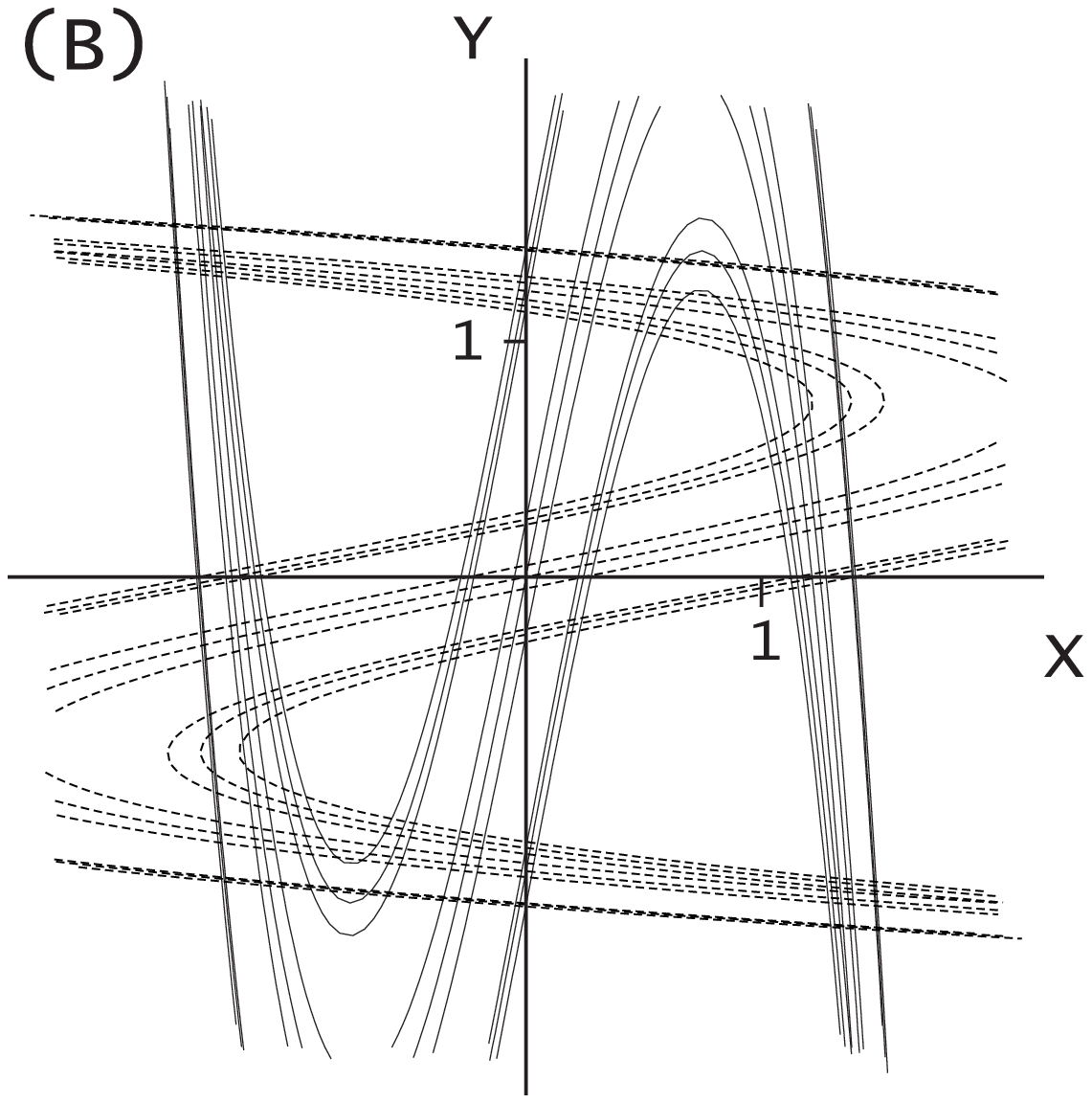}}}
{\scalebox{.6}{\includegraphics{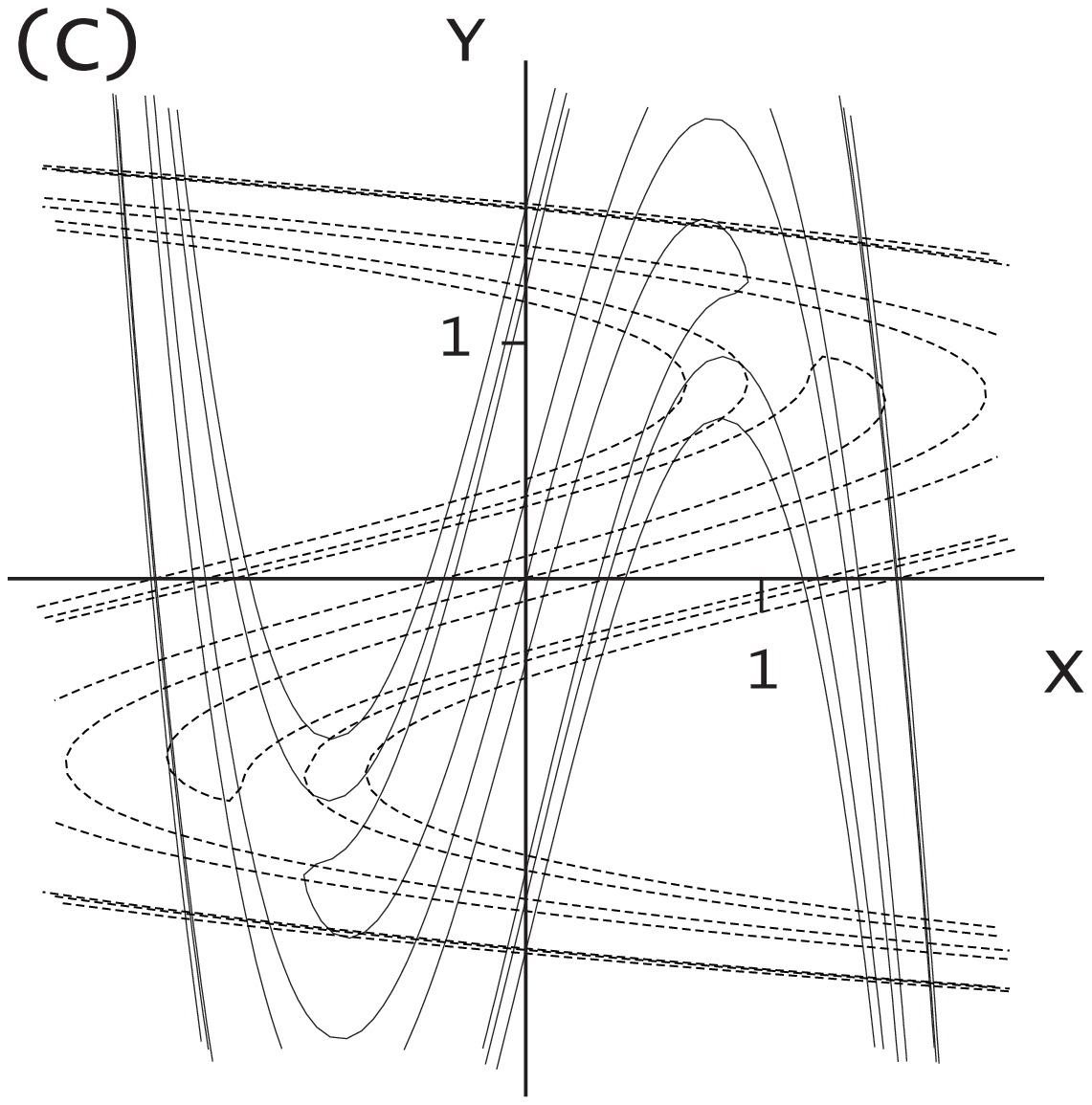}}}
{\scalebox{.6}{\includegraphics{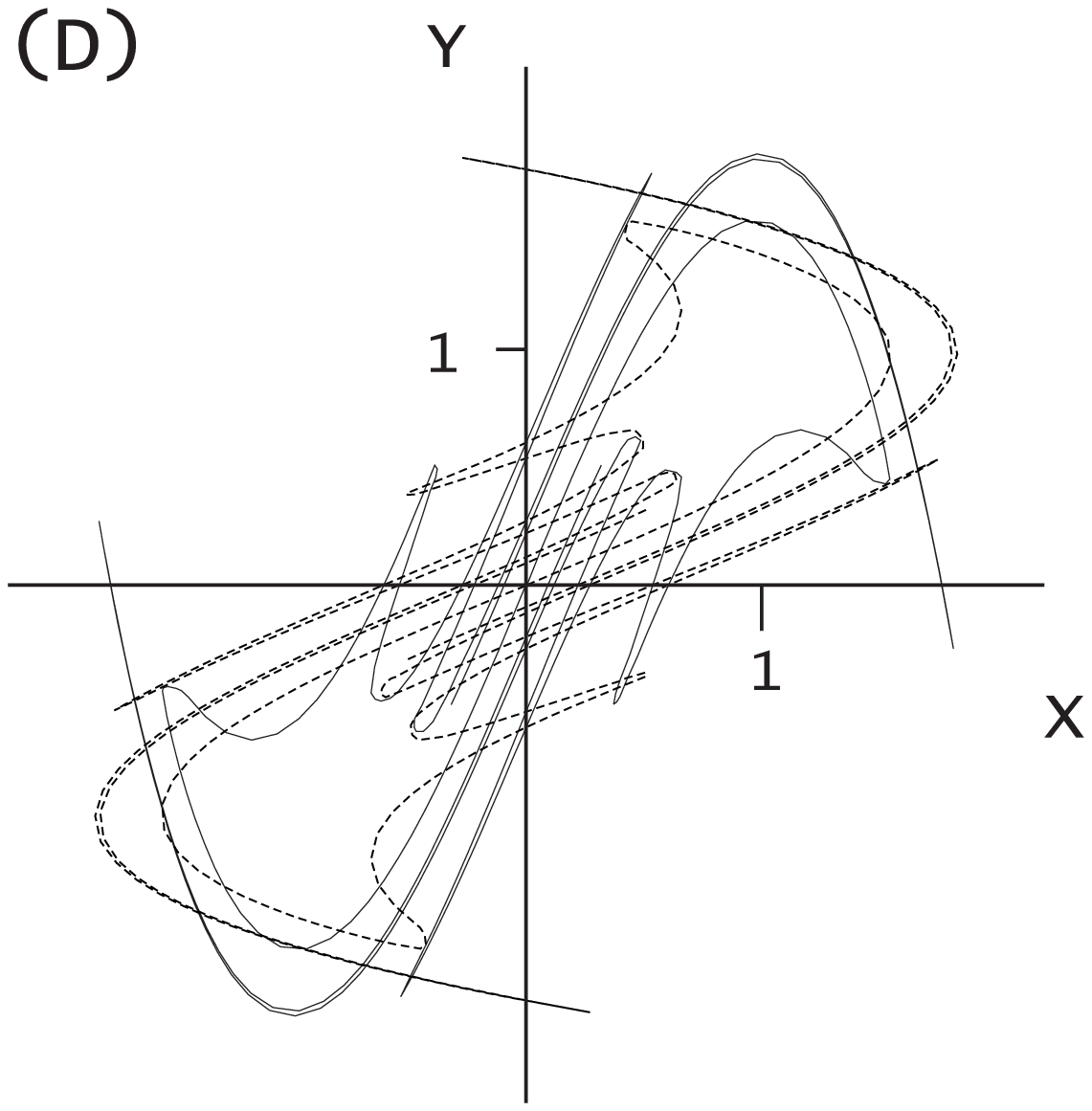}}} \caption {The stable
(dashed line) and the unstable (solid line) manifolds of the fixed
point $\mathcal O$. (a) $\alpha\approx 0.2632$, ($\omega=3.8$). In
this case $\alpha<\alpha^*$ and the structure of the intersections
of the stable and unstable manifolds is the same as in the
anticontinuous limit. (b)  $\alpha\approx 0.3125$, ($\omega=3.2$),
some homoclinic points disappeared; (c) $\alpha\approx 0.5$,
($\omega=2$); (d) $\alpha\approx 1.25$, ($\omega=0.8$).}
\label{PhaseP}
\end{figure}

\begin{conjecture}
\label{Con_Bif_1}
As $\alpha$ grows no
bifurcations of birth of new solutions occur.
\end{conjecture}

This conjecture is similar to ``No-bubbles-conjecture" of
\cite{Meiss1999} for H\'enon map. In view of this conjecture the
problem is reduced to the description of the bifurcations of death
of solutions which can be identified with their codes in the
anticontinuous limit.

Let us introduce the following notations. Let $\{\xi\}$ be a
finite word. We denote by $\{\bar{\xi}\}$ the word formed by the
symbols of $\{\xi\}$ taken in inverse order. Also we denote
$\{\tilde{\xi}\}$ the word obtained from $\{\xi\}$ by replacing
the symbols $``+"$ by $``-"$ and vice versa. We say that $\{\xi\}$
is {\it symmetric} if $\{\xi\}=\{\bar{\xi}\}$ and {\it
antisymmetric} if $\{\tilde{\xi}\}=\{\bar{\xi}\}$. Also we call
symmetric (antisymmetric) the homoclinic point coded with
symmetric (antisymmetric) word $\{\xi\}$. In the same manner we
define symmetric (antisymmetric) solutions of Eq.(\ref{AC_L}).
Then the two types of homoclinic bifurcations are generic in our
case:

\underline{\it Saddle-node} bifurcation. In a point of the
saddle-node bifurcation $\alpha=\alpha_{SN}$ two homoclinic points
which exist for $\alpha<\alpha_{SN}$, merge at
$\alpha=\alpha_{SN}$, and do not exist for $\alpha>\alpha_{SN}$.
It is clear that if two homoclinic points with codes $\{\xi_1\}$
and $\{\xi_2\}$ undergo the saddle-node bifurcation, then the
pairs of homoclinic points
$(\{\tilde{\xi_1}\},\{\tilde{\xi_2}\})$,
$(\{\bar{\xi_1},\bar{\xi_2}\})$, and
$(\{\tilde{\bar{\xi_1}}\},\{\tilde{\bar{\xi_2}}\})$ also undergo
the saddle-node bifurcation at the same $\alpha=\alpha_{SN}$.

\underline{\it Pitchfork} bifurcation. In a point of the pitchfork
bifurcation $\alpha=\alpha_{P}$ two homoclinic points with codes
$\{\xi_1\}$ and $\{\xi_2\}$ related by the symmetry
$\{\xi_1\}=\{\bar{\xi}_2\}$ or antisymmetry
$\{\xi_1\}=\{\tilde{\bar{\xi}}_2\}$ die by merging with a third
homoclinic point $\{\xi_3\}$. The code $\{\xi_3\}$ of the resulting
homoclinic point is symmetric if $\{\xi_1\}=\{\bar{\xi}_2\}$ and
antisymmetric if $\{\xi_1\}=\{\tilde{\bar{\xi}}_2\}$. This
bifurcation is generic since the dynamical system possesses
involutions.

\begin{figure}
\centerline{\includegraphics[scale=0.5]{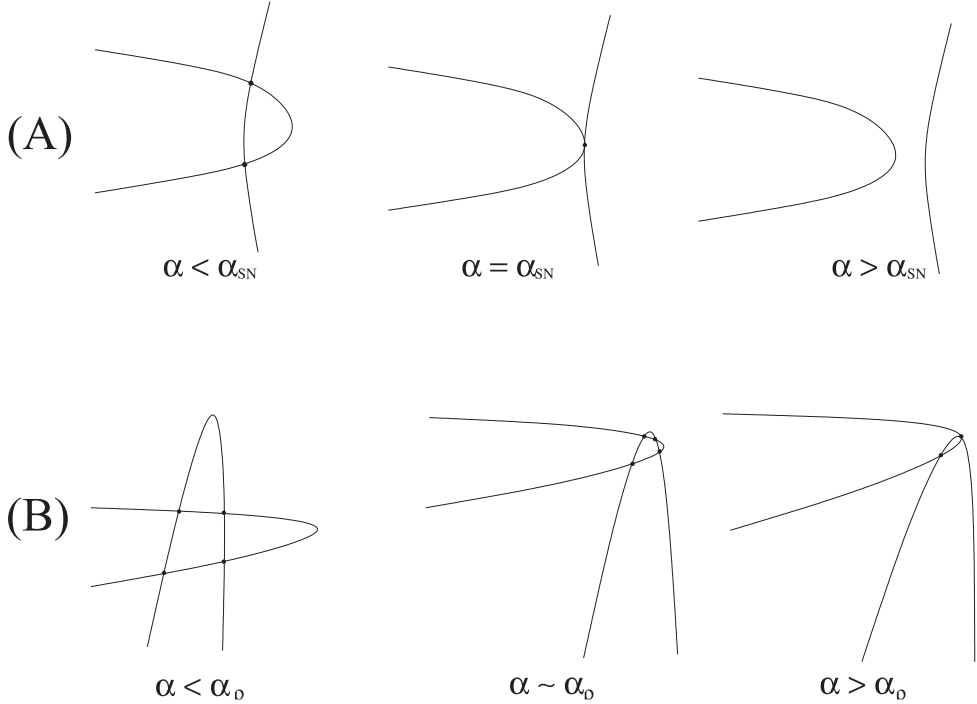}} \caption
{Saddle-node (a) and pitchfork (b) bifurcations illustrated in
terms of intersections of lobes of stable and unstable manifolds.
}\label{Bifur}
\end{figure}

The both types of bifurcations can be illustrated schematically
considering a generic intersection of the lobes of $W_s$ and $W_u$
(see Fig.\ref{Bifur}). It follows from Fig.\ref{Bifur} that a
pitchfork bifurcation is typically followed by a saddle-node
bifurcation. In this saddle-node bifurcation the symmetric
(antisymmetric) homoclinic point (the ``output" of the pichfork
bifurcation) merges with one more homoclinic point of the same
symmetry. We checked numerically (see below) that for
$\alpha>\alpha^*$ the both types of the bifurcations take place.

Following \cite{Meiss1999} we call homoclinic points $\bfa$ and
$\bfb$ {\it double neighbors} if there are no other homoclinic points
on the segments $W_u(\bfa,\bfb)$ and $W_s(\bfa,\bfb)$. Homoclinic
points cannot undergo bifurcation unless they are double
neighbors. This observation implies that

\begin{lemma}
\label{Lemma_Bif_1}
The codes of bifurcating
homoclinic points (solutions) must start and end with the same
symbol.
\end{lemma}
{\it Proof.} This statement immediately follows from the fact that
the homoclinic points with codes starting with $``+"$ lie on the
branch $W_u^+$ whereas ones with codes starting with $``-"$ belong
to $W_u^-$. So, the homoclinic points with different first symbol
of the code belong to different branches and cannot be neighbors.
The same arguments can be used for the last symbol of the code
$\blacksquare$.

Let us now turn to the analysis, what happens on the boundary of
the interval  $0<\alpha<\alpha^*$.

\begin{lemma}
\label{Lemma_Bif_2}
The value $\alpha=\alpha^*$ is
not a value of saddle-node or pitchfork bifurcation.
\end{lemma}
{\it Proof}. It follows from Theorem~\ref{Theor_Ord_1} that for
$0<\alpha<\alpha^*$ there exist infinitely  many homoclinic points
between any pair of homoclinic points which belongs to $B_s^\pm$ .
This implies that there are no double neighbors at $B_s^\pm$ for
$\alpha=\alpha^*$ $\blacksquare$.

Let $\{\xi\}$ be arbitrary finite word. Then we define the word
\begin{eqnarray*}
\{\xi\}^n\equiv\{\underbrace{\{\xi\}\{\xi\}\ldots\{\xi\}}_{\mbox{n
times}}\}
\end{eqnarray*}
and its limit case
\begin{eqnarray*}
\{\xi\}^\infty \equiv \{\ldots\{\xi\}\{\xi\}\ldots\{\xi\}\ldots\}.
\end{eqnarray*}
The results of numerical investigation allow us to formulate the
following statement:

\begin{statement} \label{Stat1}
 The value $\alpha^*\approx 0.28958$
is a limit value of a sequence of points of homoclinic saddle-node
bifurcations. Bifurcating pairs of homoclinic points have the
codes $\{+\{-+\}^n + \{+-\}^n\}$ and $\{+\{-+\}^n 0 \{+-\}^n\}$,
$n$ - positive integer. This value $\alpha^*$ also can be
understood (and found numerically in practice) as a point of
saddle-node bifurcation of the solution of (\ref{AC_L}) coded by
$\{\{-+\}^\infty + \{+-\}^\infty\}$ which is double asymptotic to
the periodic solution $\{+-\}^\infty$.
\end{statement}

\begin{figure}
\centerline{\includegraphics[scale=0.5]{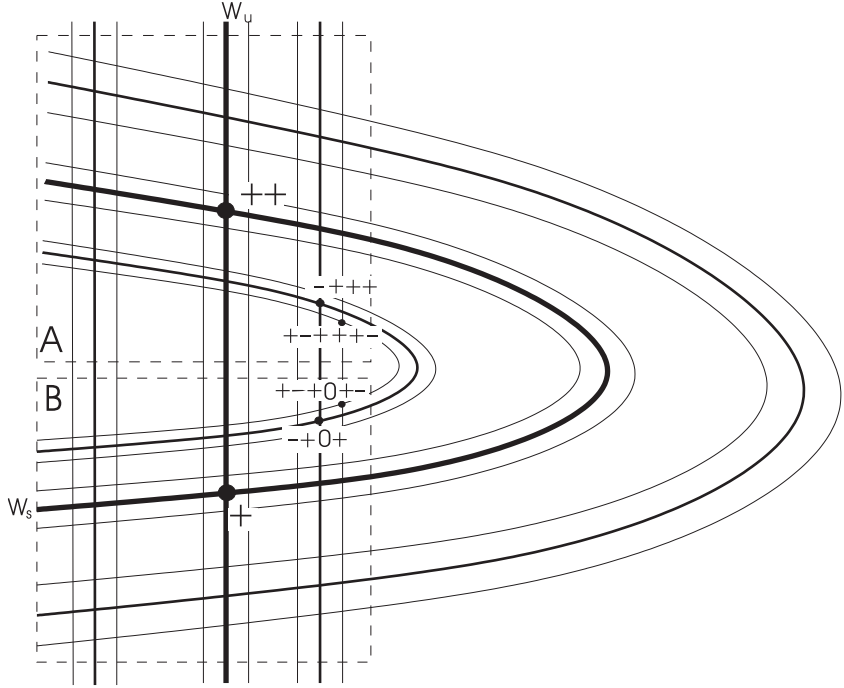}} \caption {The
parts of the stable and unstable manifolds which pass through the
area of the bifurcations, $\alpha\sim \alpha^*$. The bolder lines
correspond to the earlier passages of the stable and unstable
manifolds through the area.} \label{Area}
\end{figure}

We comment Statement~\ref{Stat1}  graphically as follows. In
Fig.~\ref{Area} there are plotted the parts of stable and unstable
manifolds which pass through the area of the bifurcations. There
are two rectangles marked as (A) and (B) in Fig.~\ref{Area}. The
coding of homoclinic points within these rectangles can be
understood from Fig.\ref{Grid}. Specifically, in the case of the
rectangle (A) one should take $\xi=\{++\}$ and reflect
Fig.~\ref{Grid} with respect to central vertical line; in the case
of the rectangle (B) one should take $\xi=\{+0\}$ and rotate the
Fig.~\ref{Grid} by $180$ degrees. As $\alpha$ grows the points
from the right lower corner of (A) tend to the corresponding
points from right upper corner of (B). Up to the lines shown in
Fig.~\ref{Grid} this corresponds to the saddle-node bifurcation of
homoclinic points with codes $\{+-+++-\}$ and $\{+-+0+-\}$.
Completing the picture of Fig.~\ref{Grid} i.e. adding  more
horizontal and vertical lines we found other codes of bifurcating
homoclinic points:
\begin{equation}
\begin{array}{ll}
\{-+-+++-+\} &\Longleftrightarrow \{-+-+0+-+\},\\
\{+-+-+++-+-\} &\Longleftrightarrow \{+-+-+0+-+-\},\\
\{-+-+-+++-+-+\} &\Longleftrightarrow \{-+-+-+0+-+-+\},\\
\vdots& \\
\end{array}
\label{BifFinSeq}
\end{equation}
Bifurcations of words with greater lengths occur at smaller values
of $\alpha$ than the ones of words with smaller lengths of code.
The codes of the bifurcating homoclinic points approach the
infinite sequences
\begin{equation}
\{\{-+\}^\infty + \{+-\}^\infty\}\quad  \mbox{and} \quad
\{\{-+\}^\infty 0\{+-\}^\infty\}. \label{BifInfSeq}
\end{equation}
These sequences correspond to the orbits homoclinic to periodic
orbit with the code $\{+-\}^\infty$. The bifurcations of these
orbits can be found by studying stable and unstable manifolds of
the fixed point $\{+\}^\infty$ of the squared map $T^2$. This
analysis shows that (\ref{BifInfSeq}) undergo saddle-node
bifurcation at $\alpha^*\approx 0.28958$. Also it was confirmed
numerically that the points of the bifurcations of pairs
(\ref{BifFinSeq}) approach this value as the length of the
corresponding codes grows.

Let us now consider in more detail the bifurcations of these
solutions which occur for $\alpha>\alpha^*$. The complete
description of the bifurcations for the solutions with codes of
length smaller than 4 are shown in Table \ref{TabBif} which was
compiled studying directly the transformation of the stable and
unstable manifolds $W_s$ and $W_u$ as $\alpha$ grows. In this
Table we show also the bifurcations related to them, for example,
pitchfork bifurcations which involves also solutions with codes of
length 5 and 6.  It worth noting that the bifurcating solutions
may have different lengths of the codes.
\begin{table}
\begin{tabular}{ccc}
$\alpha$ & 
type & codes \\
\hline
$0.324$ &
SN &
\begin{tabular}{cc}
$\{---+\}\Leftrightarrow \{-0-+\}$; &
$\{+++-\}\Leftrightarrow \{+0+-\}$; \\
$\{-+++\}\Leftrightarrow \{-+0+\}$; & \{$+---\}\Leftrightarrow
\{+-0-\}$;
\end{tabular}
\\ \hline
$0.374$ & 
SN &
\begin{tabular}{cc}
$\{+++\}\Leftrightarrow \{+0+\}$; & $\{---\}\Leftrightarrow
\{-0-\}$
\end{tabular}
\\ \hline
$0.485$ & 
P &
\begin{tabular}{c}
$\{+0++\}\Longrightarrow \{++++\}\Longleftarrow \{++0+\}$; \\
$\{-0--\}\Longrightarrow \{----\}\Longleftarrow \{--0-\}$;
\end{tabular}
\\
$0.609$ & 
SN &
\begin{tabular}{cc}
$\{++++\}\Leftrightarrow \{+00+\}$; $\{----\}\Leftrightarrow
\{-00-\}$;
\end{tabular}
\\ \hline
$0.745$ & 
P &
\begin{tabular}{c}
$\{++-+-\}\Longrightarrow \{+-+-\}\Longleftarrow \{+-+--\}$; \\
$\{--+-+\}\Longrightarrow \{-+-+\}\Longleftarrow \{-+-++\}$;
\end{tabular}
\\
$0.745+\Delta$ & 
SN &
\begin{tabular}{cc}
$\{++-+--\}\Leftrightarrow \{+-+-\}$;  $\{--+-++\}\Leftrightarrow
\{-+-+\}$;
\end{tabular}
\\ \hline
$0.747$ & 
P &
\begin{tabular}{c}
$\{++-+\}\Longrightarrow \{+-+\}\Longleftarrow \{+-++\}$; \\
$\{--+-\}\Longrightarrow \{-+-\}\Longleftarrow \{-+--\}$;
\end{tabular}
\\
$0.747+\Delta$ & 
SN &
\begin{tabular}{cc}
$\{+-+\}\Leftrightarrow \{++-++\}$; $\{-+-\}\Leftrightarrow
\{--+--\}$;
\end{tabular}
\\ \hline
$0.762$ & 
SN &
\begin{tabular}{cc}
$\{+0-+\}\Leftrightarrow \{+0-++\}$;
$\{-0+-\}\Leftrightarrow \{-0+--\}$; \\
$\{+-0+\}\Leftrightarrow \{++-0+\}$;
$\{-+0-\}\Leftrightarrow\{--+0-\}$
\end{tabular}
\\ \hline
$0.772$ & 
P &
\begin{tabular}{c}
$\{++-\}\Longrightarrow \{+-\}\Longleftarrow \{+--\}$; \\
$\{--+\}\Longrightarrow \{-+\}\Longleftarrow \{-++\}$;
\end{tabular}
\\
$0.772+\Delta$ & 
SN &
\begin{tabular}{cc}
$\{-+\}\Leftrightarrow \{--++\}$; $\{+-\}\Leftrightarrow \{++--\}$
\end{tabular}
\\ \hline
$0.779$ & 
P &
\begin{tabular}{c}
$\{++--+\}\Longrightarrow \{++--++\}\Longleftarrow \{+--++\}$; \\
$\{--++-\}\Longrightarrow \{--++--\}\Longleftarrow \{-++--\}$;
\end{tabular}
\\
$0.779+\Delta$ & 
SN &
\begin{tabular}{cc}
$\{+--+\}\Leftrightarrow \{++--++\}$;  $\{-++-\}\Leftrightarrow
\{--++--\}$;
\end{tabular}
\\ \hline
$1.091$ & 
P &
\begin{tabular}{c}
$\{++0-\}\Longrightarrow \{+0-\}\Longleftarrow \{+0--\}$; \\
$\{--0+\}\Longrightarrow \{-0+\}\Longleftarrow \{-0++\}$;
\end{tabular}
\\
$1.091+\Delta$ & 
SN &
\begin{tabular}{cc}
$\{+0-\}\Leftrightarrow \{++0--\}$; $\{-0+\}\Leftrightarrow
\{--0++\}$;
\end{tabular}
\\ \hline
$1.290$ & 
P &
\begin{tabular}{c}
$\{++00-\}\Longrightarrow \{+00-\}\Longleftarrow \{+00--\}$; \\
$\{--00+\}\Longrightarrow \{-00+\}\Longleftarrow \{-00++\}$;
\end{tabular}
\\
$1.290+\Delta$ & 
SN &
\begin{tabular}{cc}
$\{+00-\}\Leftrightarrow \{++00--\}$;  $\{-00+\}\Leftrightarrow
\{--00++\}$.
\end{tabular}
\\ \hline\
\end{tabular}
\caption{Bifurcations of solutions with the code length smaller
than 4 and related to them. By the symbol $\Delta$ we denote the
difference between the values of $\alpha$ for pitchfork and
subsequent saddle-node bifurcations when they are too close to be
distinguished by the numerics.} \label{TabBif}
\end{table}

In order to study the bifurcations of solutions with greater
lengths of the code we used the following approach. First, using
the Newton method it is possible to follow numerically the
continuation by $\alpha$ of any localized solution of
Eq.(\ref{AC_L}) coded with $\{\xi\}$ from the anticontinuous
limit. This can be done until some final point
$\alpha=\alpha_{\{\xi\}}$ where the determinant of Newton matrix
vanishes. Let at the point $\alpha=\alpha_{\{\xi\}}$ the
amplitudes of the sites obtained by continuation of the solution
with code $\{\xi\}$ be $v_n^{\{\xi\}}(\alpha_{\{\xi\}})$, $n\in
\mathbb{Z}$. If these final amplitudes are close to each other for
two different codes $\{\xi_1\}$ and $\{\xi_2\}$ and at the same time
$\alpha_{\{\xi_1\}}$ is close to $\alpha_{\{\xi_2\}}$ then one can
expect that the solutions with these codes bifurcate.

\begin{figure}
\centerline{\includegraphics[scale=0.35]{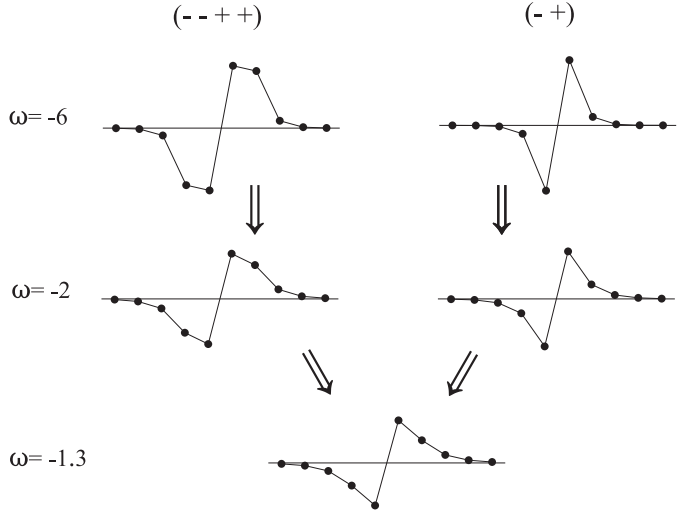}} \caption
{The saddle-node bifurcation of the solutions with the codes
$\{-+\}$ and $\{--++\}$. The point of the bifurcation is
$\alpha_{SN}\approx 0.772$ ($\omega\approx -1.295$).} \label{SF}
\end{figure}\begin{figure}
\centerline{\includegraphics[scale=0.35]{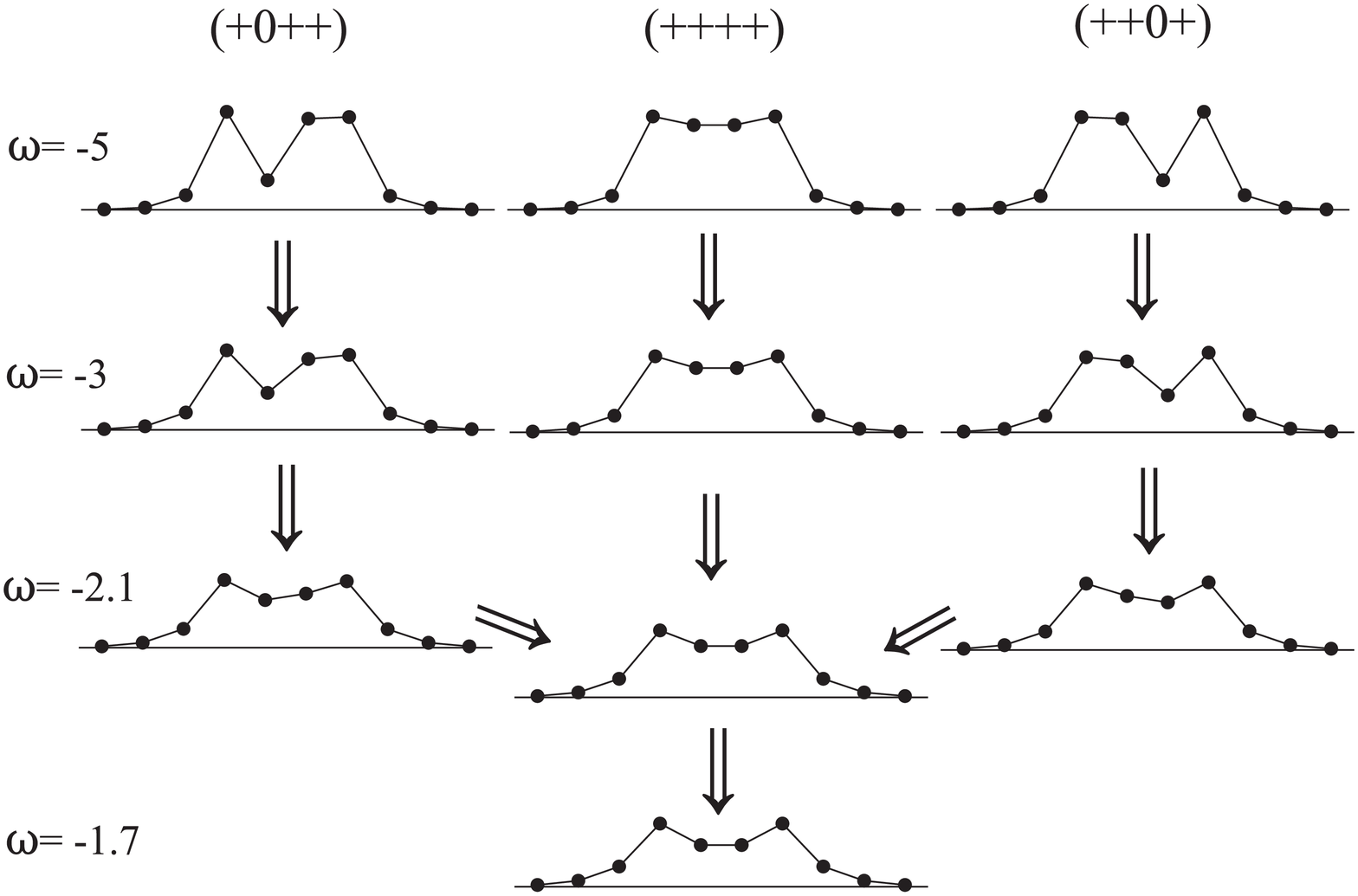}} \caption {A
pitchfork  bifurcation. Two asymmetric solution with the codes
$\{+0++\}$ and $\{++0+\}$ merge with the solution with the code
$\{++++\}$. The point of the bifurcation is $\alpha_{P}\approx
0.485$ ($\omega\approx -2.062$).} \label{PitchF}
\end{figure}

We fulfilled this continuation procedure for the solutions of
Eq.(\ref{AC_L}) which have the codes of lengths equal or smaller
than 9. Owing to the symmetries we restrict ourselves by the codes
which end up by $``+"$, i.e. $3^8=6561$ codes. Using these
continuation results we created a database. This database includes
the codes $\{\xi\}$ of the solution, the values of $\alpha_{\xi}$
and the values $\{v_n^{\{\xi\}}(\alpha_{\{\xi\}})\}$ for large
enough number of sites (basically we took 40 sites). The search
within this database allows by an introduced code of solution to
find a code of another solution which bifurcates with it. The
criterion for codes $\{\xi_1\}$ and $\{\xi_2\}$ to be the codes of
bifurcating solutions was taken as follows
\begin{equation}
\min_{|m|<10} (\sum_n
\left(v_{n-m}^{\{\xi_1\}}(\alpha_{\{\xi_1\}})-
v_n^{\{\xi_2\}}(\alpha_{\{\xi_2\}})\right)^2 <\epsilon.
\label{Crit}
\end{equation}
Here $\epsilon$ is some small positive number (we took
$\epsilon=0.1$).

We test this method using the values of bifurcation points given
in Table \ref{TabBif} and in all the cases the discrepancy was
about $10^{-3}$ which corresponds to the step in the continuation
procedure. In general, the following situations can be
distinguished:
\begin{itemize}
\item[-]
If a solution dies in a saddle-node bifurcation {\it which is not
the second bifurcation in the pair ``pitchfork - saddle-node"} the
search in the database gives one code for the solution which
bifurcate with it.
\item[-]
If a solution dies in a pitchfork bifurcation the search in the
database typically gives 3 corresponding codes. The reason is that
if the initial segment of stable (unstable) manifold is large
enough, its lobes become very narrow, so the points of pitchfork
bifurcation can be hardly distinguished from the points of
subsequent saddle-node bifurcation (in the Table \ref{TabBif} we
denote the distance between these close points of bifurcations by
the symbol $\Delta$; we should stress that $\Delta$ is not a fixed
value but any value which is below the accuracy of the method).
So, the search in the database in this case gives two codes of
solutions which participate in the pitchfork bifurcation and one
more code for the solution which dies in subsequent saddle-node
bifurcation.
\item[-]
By the same reason, if a  solution dies in a saddle-node
bifurcation which is the second bifurcation in the pair
``pitchfork - saddle-node" the search in the database typically
gives 3 codes.
\end{itemize}

\begin{figure}
\includegraphics[scale=0.5]{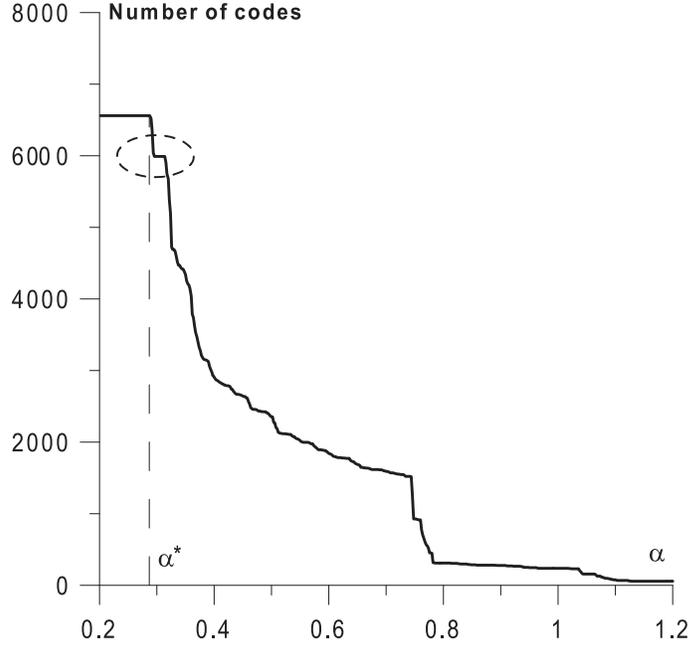}
\caption{Number of coexisting solutions with codes smaller or
equal to 9, $N(\alpha)$, versus $\alpha$.} \label{BifCurve}
\end{figure}

Also the database allows to calculate the number $N(\alpha)$ of
solutions with codes of length smaller or equal to 9 which coexist
for a given value of $\alpha$. The plot of $N(\alpha)$ versus
$\alpha$ is shown in Fig. \ref{BifCurve}.
The following comments about the bifurcations for
$\alpha>\alpha^*$ are in order.

(a) There exist intervals of $\alpha$ where no bifurcations
occur. One of them is marked in Fig.\ref{BifCurve} with ellipse.
Its boundaries are $\alpha_1\approx 0.295$ and $\alpha_2\approx
0.314$.

(b) On the interval $(\alpha^*,\alpha_1)$ the first sequence of
bifurcations occurs. At this interval there die all solutions
whose codes contain the sequences of symbols
\begin{equation}
-+0+-,\quad -+++-, \quad +-0-+, \quad +---+. \label{ProhSeq}
\end{equation}
This fact is confirmed by the analysis of intersections of stable
and unstable manifolds $W_s$ and $W_u$. The typical bifurcations
here are
\begin{eqnarray*}
\{\{\eta_1\}-+++-\{\eta_2\}\} \Longleftrightarrow
\{\{\eta_1\}-+0+-\{\eta_2\}\};\\
\{\{\eta_1\}+---+\{\eta_2\}\} \Longleftrightarrow
\{\{\eta_1\}+-0-+\{\eta_2\}\}.
\end{eqnarray*}
The end value $\alpha_1$ corresponds to the bifurcations
\begin{eqnarray*}
\{\{+-\}^\infty -+++-\{-+\}^\infty\} \Longleftrightarrow
\{\{+-\}^\infty -+0+-\{-+\}^\infty\};\\
\{\{-+\}^\infty +---+\{+-\}^\infty\} \Longleftrightarrow
\{\{-+\}^\infty +-0-+\{+-\}^\infty\}
\end{eqnarray*}
of the orbits homoclinic to periodic orbit $\{+-\}^\infty$. This
situation is similar to one found for H\'enon map
\cite{DavMcKaySan}. Following \cite{DavMcKaySan} we suppose that
for $\alpha_1<\alpha<\alpha_2$ the localized solutions of
Eq.(\ref{AC_L}) can have any code except ones which includes the
words (\ref{ProhSeq}).

(c) One more sequence of bifurcations starts with the value
$\alpha_2$ which corresponds to the bifurcations
\begin{eqnarray*}
\{\{-+\}^\infty 0+0\{-+\}^\infty\} \Longleftrightarrow
\{\{-+\}^\infty ++0\{-+\}^\infty\};\\
\{\{+-\}^\infty 0+0\{+-\}^\infty\} \Longleftrightarrow
\{\{+-\}^\infty 0++\{+-\}^\infty\};\\
\{\{+-\}^\infty 0-0\{+-\}^\infty\} \Longleftrightarrow
\{\{+-\}^\infty --0\{+-\}^\infty\};\\
\{\{-+\}^\infty 0-0\{-+\}^\infty\} \Longleftrightarrow
\{\{-+\}^\infty 0--\{-+\}^\infty\}.
\end{eqnarray*}
These bifurcations affect the solutions which codes contain the
sequence of symbols $``-+0+0"$ and other related to it due to
symmetries.  At the interval $(\alpha^*,0.341)$ all these
solutions die. The typical bifurcation is
\begin{eqnarray*}
\{\{\eta_1\}-+++0\{\eta_2\}\} \Longleftrightarrow
\{\{\eta_1\}-+0+0\{\eta_2\}\} \\
\end{eqnarray*}
and other related to it by symmetries.

(d) For $\alpha>1.21$ all the solutions in our database have the
codes which are composed by the words $``+"$, $``-"$,$``++"$,
$``--"$ separated by groups of the symbols $``0"$. So, they can be
regarded as bound states of Page and Sievers-Takeno modes. A study
of such multipulse solutions for $\alpha$ large can be found in
\cite{Kevrek}.

Since $\omega=-1/\alpha$ all these results can be rewritten  in
terms of Eq.(\ref{R_DNLS}). In particular one obtains that ACL
coding holds for $\omega<\omega^*\approx -3.4533$ and the point
$\omega^*$ is the point of accumulation of saddle-node
bifurcations of solutions of Eq.(\ref{R_DNLS}).

\section{Conclusion} \label{Concl}

To summarize, the main results of the paper are:

(a) For the equation Eq.(\ref{R_DNLS}) we used the coding of all
localized solutions (\ref{Local}) by the words which consist of
the symbols ``$-$", ``0" and ``+". This coding comes from
anticontinuous limit (ACL-coding). We calculate the boundary of
the interval of parameter $\omega$, $\omega^*\approx -3.4533$ for
which this coding is valid.

(b) In order to do this we analysed the problem from the viewpoint
of dynamical systems. Since the localized solutions of this
discrete equation corresponds to homoclinic points of zero fixed
hyperbolic point $\mathcal O$ of associated map $T$ we study the
mutual position of stable and unstable manifolds of $\mathcal O$.
A simple rule was found for the ordering of the homoclinic points
on these manifolds when anticontinuous approximation is valid.

(c) Based on this we studied the bifurcations of localized
solutions of Eq.(\ref{R_DNLS}). They correspond to homoclinic
bifurcations for the map $T$. It was shown that the end point
$\omega^*$ of the interval where the ACL-coding is applicable is
the point of accumulation of saddle-node homoclinic bifurcations.
Then we studied the bifurcations which occur for
$\omega>\omega^*$. We found an interval of $\omega$ where no
bifurcations occur and gave a complete list of bifurcations for
the solutions whose coding sequences have lengths smaller or equal
to 4.

It worth to compare our results with the results of \cite{Bount2}
and \cite{Bount3}. In both these papers Eq.(\ref{R_DNLS}) is
rewritten in the form
\begin{equation}
A_{n+1}+A_{n-1}+CA_n=3A_n^3. \label{Bount}
\end{equation}
The authors of \cite{Bount2} calculated numerically the values of
$C$ for which there exist symmetric localized solutions keeping
fixed two internal parameter corresponding to the asymptotic of
the solution for $n$ large (specifically, $\epsilon$ and $N$). The
values of these parameters were chosen for reasons of numerical
convenience. The coding of solutions introduced in \cite{Bount2},
in fact, corresponds to ACL coding. Applying our results one can
found that for $C>C^*\approx 5.4530$ all solutions of
(\ref{Bount}) can be coded by $\LL$ sequences. However, {\it all}
values of $C$ given in resulting Table 1 of \cite{Bount2} exceed
this threshold. This explains why the number of found solutions is
equal to one predicted by the analysis of number of possible
coding sequences (which, generically, does not take place).
Changing the internal parameters of the method one can make the
value of $C$ smaller than the critical value $C^*$ and the
situation immediately becomes much more complex. We note also that
the illustrative example of \cite{Bount3} related to
Eq.(\ref{Bount}) corresponds to $C=8$ which also exceeds the
threshold value $C^*$.

To complete the picture, the information about the existence of
discrete brea\-thers should be accompanied with the results on
their stability. This part of the work will be presented
elsewhere. Here we note that the analysis of bifurcations of
discrete breathers proved to be quite useful  for the study of
their stability. We sketch the relation between  these topics as
follows. In the linear approximation the stability problem can be
reduced to the eigenvalue problem for non-symmetric infinite
matrix $L$. This matrix is the product of two infinite symmetric
matrices, $L_-$ and $L_+$, where $L_+$ correspond to linearization
of Eq.(\ref{R_DNLS}) (Newton matrix) \cite{Spatschek96}. An
information about spectral properties of $L_-$ and $L_+$ can be
obtained in the anticontinuous limit. Varying $\omega$ one
observes that at the point of homoclinic bifurcation
$\omega=\tilde{\omega}$ (and not in other points) one of the
eigenvalues of $L_+$ vanishes. From the other side one can show
that the number of positive eigenvalues of $L_-$ does not change
when $\omega$ varies. These facts, together with the relation
between the numbers of positive eigenvalues of $L_-$, $L_+$ and
$L$ \cite{Jones,Gril} give a basis for the analysis of stability
of discrete breathers.

Authors are grateful to Prof. P.Kevrekidis for reading the
manuscript, pointing useful references out and many important
comments and to Prof. C. Eilbeck for his illuminating
explanations.  G.A. thanks the financial support
from Spanish Ministry of Education, Culture and Sport through a
sabbatical program SAB2000-0340. The work of V.A.B. has been supported by the FCT
fellowship SFRH/BPD/5632/2001. V.V.K. acknowledges support from
the Programme ``Human Potential-Research Training Networks",
contract No. HPRN-CT-2000-00158.

\appendix
\section{The continuation from anticontinuous limit} \label{App0}

Let us denote
\begin{equation}
{\bf F}({\bf v},\alpha )\equiv \mbox{col} (\alpha
\Delta_{2}v_{n}-v_{n}+v_{n}^{3}),\quad n\in \mathbb{Z}. \label{F}
\end{equation}
Then
\begin{equation}
\frac{d{\bf v}}{d\alpha }=-\left[ \frac{\partial {\bf F}({\bf v},\alpha )}{%
\partial {\bf v}}\right] ^{-1}\frac{\partial {\bf F} ({\bf v},\alpha )}{\partial
\alpha },  \nonumber
\end{equation}
where
\begin{eqnarray*}
\frac{\partial {\bf F}({\bf v},\alpha )}{\partial \alpha }={\rm
col}\left(\ldots,\Delta_{2}v_{n-1},\Delta_{2}v_{n},\Delta_{2}v_{n+1},\ldots\right)
\end{eqnarray*}
and
\begin{equation}
D_v{\bf F}\equiv\frac{\partial {\bf F}({\bf v},\alpha )}{\partial {\bf
v}}=\left(
\begin{array}{cccc}
\ddots  & \cdots  & \cdots  & 0 \\
\alpha  & (-2\alpha -1+3v_{n-1}^{2}) & \alpha  & \cdots  \\
\cdots  & \alpha  & (-2\alpha -1+3v_{n}^{2}) & \alpha  \\
0 & \cdots  & \cdots  & \ddots
\end{array}
\right) \label{Matrix}
\end{equation}
are an infinite column-vector and a three-diagonal infinite matrix
correspondingly. The continuation by the parameter $\alpha$ is
possible while the operator $D_v{\bf F}$ remains invertible.
According to \cite{McKay1994} this takes place until some value
$\alpha=\alpha^*$. The paper \cite{McKay1994} offers also a method
how to estimate the  boundary $\alpha^*$.

\section{Proof of Theorem \ref{Theor_Ant_lim_2}} \label{AppA}

First, we prove the following lemma:
\begin{lemma}
\label{Theor_Ant_lim_1} For $0<\alpha<\alpha_0$, where
$\alpha_0=(3\sqrt{3}-1)/52\approx 0.0807$ the operator $D_v{\bf
F}$ (see Appendix~\ref{App0}) is invertible.
\end{lemma}


\underline{\it Proof of Lemma~\ref{Theor_Ant_lim_1}.} Let us
select the conditions when the matrix (\ref{Matrix}) has main
diagonal domination (MDD). By definition, MDD means that the
modulus of each element on the main diagonal is greater than the
sum of module of other elements of the same string i.e.
\begin{eqnarray*}
\delta_n\equiv |-2\alpha -1+3v_n^{2}|-2\alpha>0
\end{eqnarray*}
for any $n\in \mathbb{Z}$. Having MDD situation one can conclude by
the generalized Gershgorin theorem that the matrix (\ref{Matrix})
is non-degenerated. Consider the amplitude of one of the sites,
$v$, and suppose that $v\geq 0$ (the case $v\leq 0$ can be
analyzed by symmetry). The value of $v$ can fall into one of the
three intervals:

(i) $0\leq v\leq 1/\sqrt{3}$;

(ii) $1/\sqrt{3}\leq v<1$;

(iii) $v\geq1$.

\underline{The case (i)}. According to Lemma \ref{Lemma_Ant_lim_1}
\begin{equation}
v-v^3\leq 4\alpha\sqrt{1+4\alpha}. \label{CaseI1}
\end{equation}
The MDD condition is
\begin{equation}
\delta\equiv 1-3v^2 > 0. \label{CaseI2}
\end{equation}
One can check that the condition (\ref{CaseI2}) holds
automatically if $0\leq\alpha< 1/12$.

\underline{The case (ii)}. Let us denote $\epsilon\equiv 1-v$;
then $0\leq \epsilon<1-1/\sqrt{3}$. The condition
(\ref{InEqLemma1}) takes the form
\begin{equation}
\epsilon^3-3\epsilon^2+2\epsilon\leq 4\alpha\sqrt{1+4\alpha}.
\label{CaseII1}
\end{equation}
From the other side one has
\begin{eqnarray*}
\delta\equiv |-2\alpha -1+3v^2|-2\alpha\geq
3v^2-1-4\alpha=3\epsilon^2-6\epsilon+2-4\alpha
\end{eqnarray*}
and the MDD takes place if
\begin{equation}
3\epsilon^2-6\epsilon+2-4\alpha>0. \label{CaseII2}
\end{equation}
One can check that Eq.(\ref{CaseII2}) holds automatically if
Eq.(\ref{CaseII1}) is satisfied and $0\leq\alpha<\alpha_0$; here
$\alpha_0=(3\sqrt{3}-1)/52$ can be found as a solution of the
system of equations
\begin{eqnarray*}
\epsilon_0^3-3\epsilon_0^2+2\epsilon_0&=& 4\alpha_0\sqrt{1+4\alpha_0};\\
3\epsilon_0^2-6\epsilon_0+2-4\alpha_0&=&0.
\end{eqnarray*}
The simplest way to make sure of the last statement is to plot in
the plane $(\epsilon,\alpha)$ the areas corresponding to
Eq.(\ref{CaseII1}) and Eq.(\ref{CaseII2}).

\underline{The case (iii)}. In this case we denote $\epsilon\equiv
1+v$; then $\epsilon>0$. The condition (\ref{InEqLemma1}) takes
the form
\begin{equation}
\epsilon^3+3\epsilon^2+2\epsilon\leq 4\alpha\sqrt{1+4\alpha}
\label{CaseIII1}
\end{equation}
and MDD condition is
\begin{equation}
\delta\geq 3v^2-1-4\alpha=3\epsilon^2+6\epsilon+2-4\alpha>0.
\label{CaseIII2}
\end{equation}
One can check that Eq.(\ref{CaseIII1}) implies Eq.(\ref{CaseIII2})
for any $\alpha>0$ and $\epsilon>0$. Taking the smallest interval
between those found in (i)-(iii) we arrive at the result of
Theorem~\ref{Theor_Ant_lim_1} $\blacksquare$.

The proof of Theorem~\ref{Theor_Ant_lim_2} follows immediately
from Lemma \ref{Theor_Ant_lim_1} and Theorem 3.3 of
\cite{Surv_Tsir}. $\blacksquare$

\section{Some definitions from dynamical systems theory } \label{AppB}

\begin{definition}
A {\it segment} $W_s[\bfa,\bfb]$ of a stable manifold is the arc
of the manifold between points $\bfa$ and $\bfb$ which includes
these points. Correspondingly, $W_s(\bfa,\bfb)$ is the arc of the
manifold between the points $\bfa$ and $\bfb$ excluding these
points.
\end{definition}
In the same manner one can define the segments $W_s[\bfa,\bfb)$
and $W_s(\bfa,\bfb]$. The segments on the unstable manifold are
defined similarly.

\begin{definition}
An {\it initial segment} of the branch of a stable manifold,
$W_s^\pm({\mathcal O},\bfa]$, is the segment of $W_s^\pm$ which
extends from the hyperbolic fixed point $\mathcal O$  to the point
$\bfa\in W_s^\pm$.
\end{definition}
Similarly the initial segment of the unstable manifold,
$W_u^\pm({\mathcal O},\bfa]$ can be defined.
\begin{definition}
A {\it fundamental segment} of a branch of a stable manifold,
$W_s^\pm(\bfa, T^{-1}\bfa]$, $\bfa\in W_s^\pm$ is the segment of
$W_s^\pm$ between the point $\bfa$ and its $T^{-1}$-image, the
point $\bfa$ is excluded and $T^{-1}\bfa$ is included.
Correspondingly, a fundamental segment of the unstable manifold,
$W_u^\pm(\bfa, T\bfa]$, is the segment of $W_u^\pm$ between the
point $\bfa\in W_u^\pm$ (excluded) and its $T$-image (included).
\end{definition}

Some comments are in order. The maps $T$ and $T^{-1}$ transform
fundamental segments into other fundamental segments. Namely, for
any $\bfa\in W_u^\pm$ (or, correspondingly, $\bfa\in W_s^\pm$)
\begin{eqnarray*}
&&T\{W_u^\pm(\bfa, T\bfa]\}=W_u^\pm(T\bfa, T^2\bfa];\\
&&T\{W_s^\pm(\bfa, T^{-1}\bfa]\}=W_s^\pm(T\bfa, \bfa].
\end{eqnarray*}
The whole branch of a stable (unstable) manifold is the union of
all its fundamental segments
\begin{eqnarray*}
&&W_u^\pm=\bigcup_{n=-\infty}^\infty T^n\{W_u^\pm(\bfa,T\bfa]\},
\quad \mbox{for any $\bfa\in W_u^\pm$};\\
&&W_s^\pm=\bigcup_{n=-\infty}^\infty T^n\{W_s^\pm(T\bfa,\bfa]\},
\quad \mbox{for any $\bfa\in W_s^\pm$}.
\end{eqnarray*}


\begin{thebibliography}{99}
\bibitem{Surv_Tsir}  D. Hennig, G. Tsironis, {\it Phys. Reports} {\bf 307},
333 (1999).

\bibitem{Surv_Kevr}  P. G. Kevrekidis, K. \O . Rasmussen, A. R. Bishop, {\it
Int. J. Mod. Phys.} {\bf B15}, 2833 (2001).

\bibitem{Surv_Flach}  S. Flach, C. R. Willis, {\it Phys. Reports} {\bf 295},
181 (1998).

\bibitem{Eilbeck} J. Ch. Eilbeck, M. Johansson, in ``Localization and Energy Transfer in Nonlinear
Systems'', edited by L. V\'azquez, R.S. MacKay, and M.P Zorzano, (World Scientific, Singapore, 2003), p.44; 
arXiv: nlin.PS/0211049 (2002).





\bibitem{Bose} Abdullaev  F. Kh., Baizakov  B. B.,
 Darmanyan  S. A., Konotop V. V., and Salerno M.,
{\it Phys. Rev.} A {\bf 64} 043606 (2001); G.L.Alfimov, P.G.Kevrekidis,
V.V.Konotop and M.Salerno. {\it Phys.Rev.E}, {\bf 66},(4), 046608 (2002).


\bibitem{prb58} G. Kalosakas,   S. Aubry, and
G. P. Tsironis, Phys. Rev. B {\bf 58}, 3094 (1998).



\bibitem{Morg}  A.M. Morgante, M. Johansson,
G. Kopidakis et. al., {\it Physica} {\bf
D162}, 53 (2002).

\bibitem{ST}  A. J. Sievers, S. Takeno, {\it
Phys. Rev. Lett.} {\bf 61}, 970 (1988).

\bibitem{Page}  J. B. Page, {\it Phys. Rev.} {\bf B41}, 7835 (1990).

\bibitem{Spatschek96}  E. W. Laedke, O. Kluth, and K. H. Spatschek, {\it
Phys. Rev.} {\bf E54}, 4299 (1996).

\bibitem{Darm} S. Darmanyan, A. Kobyakov, and F. Lederer, {\it Sov. Phys.
JETP} {\bf 86}, 682 (1998).

\bibitem{Bount2}  J. M. Bergamin, T. Bountis, C. Jung, {\it J. Phys.A: Math.
Gen.} {\bf 33}, 8059 (2000).

\bibitem{Bount3} J. M. Bergamin, T. Bountis, M. N. Vrahatis, {\it
Nonlinearity} {\bf 15}, 1603 (2002).


\bibitem{McKay1994}  R. S. MacKay, S. Aubry, {\it Nonlinearity} {\bf 7}, 1623
(1994).

\bibitem{Meiss1999} D. Sterling, H. R. Dullin, and J. D. Meiss, {\it Physica
}  {\bf D134}, 153 (1999).

\bibitem{Meiss2000} H. R. Dullin, J. D. Meiss,  {\it Physica} {\bf D143},
265, (2000).

\bibitem{pre52} D. Hennig, N. G. Sun, H. Gabriel, and G. P. Tsironis, Phys. Rev.
 E {\bf 52}, 255 (1995);

\bibitem{pre54} D. Hennig, K. \O. Rasmussen,
 H. Gabriel, and A. B\"ulow, Phys. Rev. E {\bf 54}, 5788 (1996).





\bibitem{GeomMeth} R. W. Easton, {\it Geometric Methods for Discrete
Dynamical Systems}, Oxford University Press, 1998.

\bibitem{DavMcKaySan} M. J. Davis, R. S. MacKay, and A. Sannami, {\it Physica }
{\bf D52}, 171 (1991).

\bibitem{Kevrek} P. G. Kevrekidis, {\it Phys.Rev.} {\bf E64},
026611 (2001).

\bibitem{Jones} C. K. R. T. Jones, {\it Ergod. Theor. Dynam. Syst.} {\bf 8}, 119 (1988).

\bibitem{Gril} M. Grillakis, {\it Comm. Pure. Appl. Math.} {\bf 41}, 747 (1988).





\end{thebibliography}
\end{document}